\documentclass[a4paper,10pt,american]{report}

\usepackage[T1]{fontenc}
\usepackage[utf8]{inputenc}
\usepackage{babel}
\usepackage{a4}
\usepackage{url}
\usepackage[colorinlistoftodos]{todonotes}
\usepackage{fancyhdr}
\usepackage{graphicx}
\usepackage{float}
\usepackage{amsmath}
\usepackage{amsfonts}
\usepackage{scalefnt}
\usepackage{listings}
\usepackage{lipsum}
\usepackage[bookmarks=true,
  breaklinks=true,
  hypertexnames=true]{hyperref}
\usepackage{xspace}
\usepackage{xcolor}
\usepackage{enumerate}
\usepackage{csquotes}
\usepackage[yyyymmdd,hhmmss]{datetime}
\usepackage{natbib}[abbrvnat]

\usepackage{palatino}
\usepackage{inconsolata}

\newcommand{\comment}[1]{}

\newcommand{\eg}{e.g.\@ }

\newcommand{\nseg}{\mathbf{n}}
\newcommand{\iou}[2]{{}^{#1}[#2]}
\newcommand{\DKK}{\mathrm{DKK}}
\newcommand{\chapterauthor}[1]{{#1}\vspace{1cm}}

\title{Contract-Backed Digital Cash}
\author{S\o ren Debois \and Fritz Henglein \and Morten C.~ Nielsen \and Christian Olesen \and Gert Sylvest}
\date{Editor: Fritz Henglein. Drafted 2019-02-09, revised 2021-05-10, finalized 2021-07-14, last edits 2021-08-09, published 2022-11-19}

\begin{document}

\maketitle

\chapter*{Summary}

We present a series of essays on \emph{digital money} combined with secure transactional execution of \emph{digital contracts}.

In Chapters~\ref{exchange-theory} to \ref{digital-cash}, based on an exchange theory of money Henglein decomposes money into orthogonal aspects and identifies \emph{digital cash} as the digital equivalent of physical cash: secure, fungible, decentralized, directly controlled, private; but enhanced with qualitatively new functionality: being extremely efficiently transferable and storable and, most importantly, \emph{contract-backed}. This facilitates fully digitalized, inexpensive, guaranteed transactional execution such as atomic resource exchanges, without a multitude of intermediaries and expensive or slow semi-manual processes.  A didactic objective is decomposing properties of money and blockchain/DLT into independent functional aspects to illustrate the enormous unexplored design space, both for concrete digital money designs and not yet explored or under-utilized distributed systems techniques for implementing them.  This is aimed at disentangling discussions of digital money design (policy and governance) from potentially misleading and unnecessary preconceptions about particular database, distributed computing, cryptography or data structure techniques, such as Byzantine consensus, cryptographic hashing, Merkle trees, UTxOs, blind signatures, zero-knowledge proofs, etc.

In Chapter~\ref{chicago-plan}, Henglein and Olesen review the contemporary banking system and take a critical look at its insistence on relegating individuals and non-bank businesses to private money, which is not secure since it is subject to default risk. Is it really meaningful that ordinary individuals' and businesses' digital money is exposed to credit default risk whereas banks, the specialists in assessing credit default risk, have access to guaranteed default-free digital money?

In Chapter~\ref{blockchain}, Henglein presents a general functionality-oriented framework for blockchain and distributed ledger technology, understood as peer-to-peer platforms for securely managing ownership and exchanges of money and other resources.  Their key property is decentralized governance: operating without pre-appointed parties that have privileged access to or control of the system.  Decentralization is arguably essential for a free economy with fair competition and thus for digital money. In a centralized digital money system, third parties such as IT companies and banks operating it, gain an inherent competitive advantage since money is involved in \emph{all} economic transactions.

The final chapters discuss and illustrate the power and role of programmable (contract-backed) digital money in case studies.  

For the private sector, in Chapter~\ref{tokenization}, Sylvest describes effective tokenization of invoice debt using smart contracts on Ethereum, with stablecoins as digital money pegged to US dollars.  This corresponds to debt securitization---packaging commercial zero-coupon bonds into tradable securities\footnote{Without calling them securities}---but effectively available at much lower cost to small and medium-size enterprises.  

For the public sector, in Chapter~\ref{municipal-administration}, Debois argues the benefits of smart contracts for disbursing payments transparently and reliably in accordance with social legislation.

Finally, in Chapter~\ref{e-krone}, Nielsen and Olesen propose the establishment of a Danish E-krone with limited usage or limited size, describe a use case and argue its political and economic benefits.

Collectively, the essays illustrate the existing and, in particular, future potential of digital money when powered by smart digital contracts that effectively eliminate both counterparty risk (somebody doesn't pay or doesn't deliver) and settlement risk (a trade fails and needs to be aborted) by orders-of-magnitude faster settlement than current practice.

The essays should be of interest to public and private actors (re)considering the role of digital money with secure end-to-end digitalization in a modern society such as Denmark, which provides an efficient digital infrastructure that has an authoritative legal entity register, digital signatures, land registry, energy production data, public data hubs and more that are ready to be integrated into fully automated digital contracts that incorporate transactional digital money for secure economic exchanges.  This includes central banks who are competing with both Big Tech and people in hoodies to control the kind of digital cash that \emph{will} be provided to ordinary citizens and businesses, whether they know it or like it; banks that know that money is about economic \emph{exchanges}, not  individual electronic \emph{transfers} (which has been a solved problem since the 1960s), and that climate change/green transition, sustainability, funding of SMEs and efficient trade finance require novel, securely and transparently automated forms, analysis, monitoring and enforcement of financial contracts; regulators who may be tempted to enshrine proven---and outdated---practices and technology into conservative regulation without eying the new opportunities of incorporating high-performance privacy-preserving technology for reducing systemic risk and enforcing anti money laundering and counter terrorism funding laws made possible by contract-backed digital cash; politicians and administrators committed to efficiently, fairly and transparently providing legally compliant services to their citizens; and fintech companies ready to pounce on the possibilities of building on top of user-definable smart contracts with bonafide quasi-central-bank issued digital currency by unleashing the computational power of what would have been classified as supercomputers twenty years ago: smartphones.   

\section*{Authorship and working group on digital cash}

The chapters have been authored separately.  The authors (only) are responsible for their contents; see the chapter author(s) at the beginning of each chapter.  The chapters have been lightly copy-edited for inter-chapter cohesion.  The report has been delayed by more than two years due to adversarial circumstances encountered by the editor in the first half of 2019.  

The working group on digital cash, which operated in 2018, provided valuable discussions and direction. Its meetings were open and attended by members affiliated with the project partners and other organizations, including Michel Avital (CBS), S\o ren Debois (ITU), Boris D\"udder (DIKU), Jonas Hedman (CBS), Fritz Henglein (DIKU and Deon Digital), Kim Peiter J\o rgensen (ITU), Morten Nielsen (Aryze), Christian Olesen (DIKU), Omri Ross (DIKU), Peter Sestoft (ITU), Gert Sylvest (TradeShift Frontiers).  

\section*{Acknowledgements}

The project \emph{contract-backed digital cash} has been made possible by a grant by the Danish Innovation Network for Finance IT to Copenhagen Business School (CBS), the IT University of Copenhagen (ITU) and the Department of Computer Science at the University of Copenhagen (DIKU). It has been supported in kind by Tradeshift Frontiers, MakerDao, Aryze, and Deon Digital.  It has benefited from the invaluable practical help by Copenhagen FinTech and encouragement by the Copenhagen FinTech Lab community.

\tableofcontents

\chapter{An exchange theory of money}
\label{exchange-theory}

\chapterauthor{Fritz Henglein}

We develop a theory of money based on bilateral resource exchange as the basic economic primitive, where one economic agent gives something to another only in exchange for receiving something from that agent in return.  The challenge is how to compose such exchanges into achieving a multi-lateral exchange that optimizes resource distribution to achieve a fabled Arrow-Debreu equilibrium \citep{glicksberg1952further,arrow1954existence}.  This is analogous to figuring out how to implement multi-party consensus using pairwise message passing in distributed computing \citep{axtell2003economics}.  While taking distributed consensus as a magical primitive with instantaneous effect (as in blockchain systems and much of rational agent economics) is a wonderful and powerful primitive both in computer programming and in economic theory, implementing it using pairwise exchange between unreliable and self-interested agents, respectively network nodes, turns out to be almost arbitrarily hard to achieve.  The notion of money as a real thing pops prominently up in this implementation struggle, that is in the machine room of economics, not so much in macroeconomics.

The present account of money is completely ahistorical.  Is is an entirely made-up and simplified rationalization, but it should be familiar, if not entirely covered, by traditional tit-for-tat accounts based on coincidence of wants \citep{10.1257/000282802320189014,kiyotaki2001evil}, payments \citep{KAHN20091} or duality of economic events \citep{mccarthy83}. 
Alexander Del Mar complained that ``[a]s a rule political economists [\ldots] do not take the trouble to study the history of money; it is much easier to imagine it and to deduce the principles of this imaginary knowledge.'' \citep{del1895history} Our account would only aggravate his complaint since it is not even given by an economist, but comes from a computer science perspective where synthesizing models at will is the rule of the day.  

Nonetheless, we hope it can serve as a coherent story of money as a real thing, even and especially when entirely manufactured \emph{ex nihilo}, and as being inextricably connected to \emph{contracts}: goods now for money now, goods now for money later, money now for money later. In general, a contract is a prescriptive multi-party protocol for transferring resources and transmitting information, where a single transfer of money in isolation is almost never the full story.  The \emph{single} act of handing over physical cash or digitally transferring bank money is a solved problem; ensuring that \emph{all} transfers required by a contract happen is unfortunately not solved by that.


\section{Resource permutations by exchanges}

Let us assume we have a set of agents, where each agent has an economic resource to offer and is in need of a potentially different resource for consumption or to produce something.   
For simplicity we will also assume that each agent has a single resource to offer and a single resource she needs and that the agents collectively have enough resources to satisfy all needs.  This can be generalized mathematically to weighted resource baskets without changing the conceptual fundamentals \citep{rambaud2010algebraic,torresgarcia2020}.

If we number the agents $\nseg = \{1, \ldots, n\}$, we can describe the resources they own at the beginning by a vector $\vec R = [R_1, \ldots, R_n]$.  The desired end state is a permutation of $\vec R$, that is one-to-one function $\pi : \nseg \rightarrow \nseg$ that moves resource $R_i$ originally owned by agent $i$ to agent $\pi(i)$.  E.g.\ if the agents own $R, S, T$, after an exchange between $1$ and $2$ they own $S, R, T$, respectively: 
	\[ \begin{array}{|lll|}
	\hline 1 & 2 & 3 \\ \hline
	R & S & T \\
	S & R & T \\ \hline
	\end{array} \]

A single pairwise \emph{exchange} corresponds to a transposition in terms of permutation theory.
If an agent exchanges her resource only for a resource she ultimately needs she can only exchange with another agent who happens to need exactly what she has.  There is no way of achieving arbitrary permutations by only executing such \emph{coincidence-of-wants} fulfilling exchanges.  Every permutation is accomplishable by a sequence of exchanges, though: It \emph{necessarily} requires (some) agents to be willing to accept a resource in an exchange even though it is \emph{not} what they actually need, but is used in a future exchange for their actually needed resource.  For example, if $[S, T, R]$ represent the needs of our three agents initially owning $[R, S, T]$, there are no immediate-wants exchanges reaching the final state, but if agent $2$ is willing to accept $R$ in the first exchange, the following exchanges reach the desired end state:
\[ \begin{array}{|lll|}
\hline 1 & 2 & 3 \\ \hline
R & S & T \\
S & R & T \\ 
S & T & R \\ \hline
\end{array} \]

Note that agent $2$ transitions through ownership of $R$ on her way from initially owning $S$ to eventually owning $T$.

\section{Commodity money: Resources for exchange} 

Accepting a resource for the sole purpose of trading it away again is a delicate mind game: It is based on the agents' mutually reinforcing belief that accepting it is okay because somebody else will accept it down the line. Such a resource serves as a \emph{medium of exchange}. Intuitively, that belief is bootstrapped from the knowledge that somebody can actually use the resource (for consumption or for production of something), and it is reinforced if the resource is observably and effectively used in exchanges by others; is fungible and countable (a commodity); retains its value over time and space; and is easy to transport and store. 
Such a resource constitutes \emph{commodity money}.  

Historically, several different resources have served as commodity money more or less well, including gold, silver, barley and cows \citep{zarlenga2004lost}.  Its perceived preservation of value derives from the commodity having \emph{intrinsic value}: it can be consumed, or something useful can be produced from it.  The primary use of a commodity serving as money is for being traded multiple times, by multiple agents, rather than being consumed. It is accepted in a trade for the express purpose of being traded away again rather than consumed or used for production.  As we have seen, the need for one or few types of physical resources serving as money in exchanges can be derived from the decomposition of macroeconomically beneficial multilateral resource exchanges into a set of bilateral resource exchanges.  It requires that an agent accepts at least one more resource type in exchange for the products she produces than the ones she is interested in consuming.  The macroeconomic benefit of accepting (commodity) money, something agents accept in exchange for their products even though they cannot directly make use of the money received, does not in itself provide an incentive to a self-interested agent to accept it.   Translating the benefit into self-interest can be explained in multiple ways: by a Peter Pan theory (self-reinforcing belief that it's okay to accept it because others believe the same) or Mahagonny theory (the requirement to pay taxes using it guarantees demand for it), FOMO (other agents having more trades by accepting money), mechanism enforcement (an authority requiring accepting money as legal tender to assure macroeconomic---or just their own---benefits), etc.

As for FOMO, agents not accepting commodity money avoid the risk of being stuck with something they do not need themselves, but cannot efficiently participate in  multilateral economic exchanges that effectively transport resources through time and space to where they are most useful.  This is a competitive disadvantage compared to agents participating in multilateral exchanges by accepting commodity money, which in turn increases the competitive pressure on agents to accept commodity money.   

An attractive feature of commodity money is that it not only serves as a medium of exchange, but serves as a store of value that frees the exchange from having to take place at the same time and place: What is needed can be gotten at a later time (by storing the commodity money) and at a different place (by transporting the commodity money there).  This provides an effective way of delaying to committing to a particular need and its time and location of delivery: an agent can opportunistically trade away her resource right away and at any opportune location, without having to commit to a particular need of her own at the same time or place.

\section{Outside money} 

Commodity money such as gold derives its value from being an effective medium of exchange.  But what if interacting agents don't have any commodity money on hand?  Consider our scenario 
	\[ \begin{array}{|lll|c|}
	\hline 1 & 2 & 3 & \mbox{state} \\ \hline
	R & S & T & \mbox{initial} \\
	S & T & R & \mbox{final} \\ \hline
	\end{array} \]
of three agents interested in a permutation of their bespoke resources $R, S, T$, none of which is trusted to be commodity money.  Since the permutation is not a coincidence-of-needs permutation\footnote{A permutation with maximal cycle length 2.} they are stuck.

One solution is bringing in a fourth agent $0$ who owns commodity money $G$; see state 1 of Figure~\ref{outside-money-exchange}.  By exchanging $G$ for $R$ with agent $1$ (state 2) she injects commodity money into the system, whereupon $1$ can pay $G$ for receiving $S$ from $2$ (state 3),
then $2$ can pay $G$ to $3$ for $T$ (state 4), and eventually $3$ can pay $G$ to $0$ for $R$ (state 5). 

\begin{figure}
\[ \begin{array}{|llll|c|}
\hline 0 & 1 & 2 & 3 & \mbox{state} \\ \hline
G & R & S & T & \mbox{1} \\
R & G & S & T & \mbox{2} \\ 
R & S & G & T & \mbox{3} \\
R & S & T & G & \mbox{4} \\
G & S & T & R & \mbox{5} \\ \hline
\end{array} \]
\caption{Resource permutation facilitated by agent providing outside commodity money}
\label{outside-money-exchange}
\end{figure}

Agent $0$ starts and ends with owning $G$ and is the only agent who accepts a resource (in the first exchange) that is neither commodity money nor what she eventually wants, which, to top the peculiarity off, is just what she already has to start with.  Why would Agent 0 participate at all if she already has what she needs?  She provides \emph{outside money}, commodity money that she injects from the outside into the system of trading agents $1, 2, 3$; it unlocks their stalemate, which is due to their lack of coincidence of wants.  After receiving back $G$ in the final step, Agent $0$ can engage with another set of agents using the \emph{same} commodity money $G$.  

Note that none of the agents $1, 2, 3$ own $G$ at the beginning or at the end of the interaction, and agent $0$ owns the same at the beginning and at the end, but, magically, the $G$ of agent $0$ provided the lubricant for facilitating the multiway exchange between $1, 2$ and $3$.  

In this fashion a single ``special'' agent owning commodity money can facilitate arbitrary multiway permutations where (ordinary) agents only accept the resource they need for consumption or production or commodity money.  

There are still some problems:
\begin{enumerate}
	\item There might be too little money.  The permutation is made possible by a single $G$, but this is extremely slow: Only one exchange is facilitated at a time.  For $n$ agents $1, 2, \ldots, n$ we need $n + 2$ rounds of exchanges, where only one exchange is performed during each round.  The average production activity, measured in units of $G$, per agent per round is $O(\frac{1}{n} \cdot G)$ since only one agent receives the resource they need for production, even though there is $O(G)$ production capacity in a round.  If $0$ had $n$ units of $G$, all exchanges could be completed in $2$ rounds instead of $n$ rounds: First, in parallel, each agent $1, \ldots, n$ sells their resource for $G$ to agent $0$; then, also in parallel, each agent $1, 2, \ldots, n$ buys the resource they need for $G$ from $0$.

As more agents with production capacity arrive, standing in line for a fixed number of $G$ increases their demand and thus the value of $G$ (in terms of the amount of resources bid for it)---we get deflation.
	\item Conversely, there might be too much money.  If production capacity decreases and/or more $G$ become available, the value of the $G$  is depressed---we get inflation.
	\item The commodity money $G$ brought in by designated agent $0$ might be used up by some agent and not returned to Agent 0, preventing further exchanges.  In other words, the uses as a commodity and as money compete with each other.
	\item The same commodity is used as money in multiple exchanges, which may take place at different locations and at different times.  This requires the commodity to be securely storable and efficiently transportable. A resource suitable for consumption is prone to be perishable, however; and a production resource is likely to be bulky and expensive to transport.
\end{enumerate}

The problem with deflation is that it has a self-reinforcing depressing effect on the economic activity it is supposed to make possible. An owner of deflationary commodity money is incentivized to hold on to it since, at any given point in time, it is likely to be worth even more (in terms of other resources) in the future, leading her to ultimately \emph{never} actually trade it for anything, which negates its economic function as medium of exchange. 

The problem with inflation is that the decline in value (in terms of other resources) limits the time the money retains its exchange value, thus also limiting its transportation to other places.  It incentivizes its holder to \emph{immediately} exchange it for a consumption or production resource.  This limits its economic function as mediator that breaks the coincidence-of-wants conundrum: ultimately one might as well engage in a barter without the money as an intermediary, and we are back at the problem of gridlocking the economy by limiting it to coincidence-of-wants exchanges.

The problem with commodity money being both commodity and money is that it may have an incidental inflationary or deflationary effect.  Considering gold as commodity money, if there is a sudden fad to wear expensive watches or buy new electronic devices or get gold crowns, this may suddenly reduce the amount of gold available as medium of exchange and lead to deflationary economic depression; once the fad is over and replaced by wearing charms made from biodegradable materials, the sudden availability of recycled gold may lead to inflationary anxiety over losing its value.  Paradoxically, while the trust in commodity money derives from its usability as commodity its predictable value to the economy as a facilitator of arbitrary resource permutations derives from it actually \emph{not} being used in production or consumption.
 
\section{Representative money}

The problem with commodity money being bulky or perishable can be addressed by replacing the commodity proper by a transferable promise to deliver it upon request; this is \emph{representative money}.  A prominent form of representative money is the US Dollar between 1900 and 1971, which represented an IOU of a certain amount of gold.

The basic idea is transferring a bearer certificate of ownership instead of transferring the actual commodity in an exchange.   The commodity can be kept at a safe storage place under the guard of a designated administrator who issues such bearer certificates of ownership.  The certificates express \emph{ownership} of a commodity by an agent and its separate \emph{possession} by the administrator.  They serve as proxies for the commodity in exchanges. 

Let us write $\iou{k}{R}$ for a transferable \emph{promissory note (``IOU'')} issued by agent $k$, in which $k$ promises to turn over $R$ to any agent who owns the note in exchange for the note itself.  The note itself is now a resource that can be transferred.  If the resource is commodity money $G$ and the issuer is widely trusted to satisfy its obligation, the note constitutes \emph{representative money}.  We write $-\iou{k}{R}$ for $k$'s obligation to deliver on the promise.  This is a \emph{liability}, a \emph{negative} resource.  Not ``owning'' it is preferable to owning it.  

The magic of promises as resources is that we can generate them out of nothing.  An agent can split $0$ (nothing) into a transferable, positive promise to turn something over and the complementary negative obligation of turning it over on demand: 
$$ 0 = \iou{k}{R} + (-\iou{k}{R})$$
or simply $0 = \iou{k}{R} -\iou{k}{R}$.
For now, we don't allow negative resources to be transferable: It is a liability that an agent ``owns'' and that can only be canceled if she delivers on it by turning over the actual resource in exchange for the note---or by receiving the note back in exchange for something else.

Recall the exchanges in Figure~\ref{outside-money-exchange}, where Agent $0$ injects liquidity in the form of $G$ to facilitate the exchanges by the other agents $1, 2, 3$.  Since $G$ is not consumed, but used for exchanges only, she can replace commodity money $G$ by representative money $\iou{0}{G}$ and keep the $G$ in storage instead; see Figure~\ref{outside-money-exchange-iou}.

\begin{figure}
\[ \begin{array}{|llll|}
\hline 0 & 1 & 2 & 3 \\ \hline
G & R & S & T \\
G + \iou{0}{G} - \iou{0}{G} & R & S & T \\
G + R - \iou{0}{G} & \iou{0}{G} & S & T \\ 
G + R - \iou{0}{G} & S & \iou{0}{G} & T \\
G + R - \iou{0}{G} & S & T & \iou{0}{G} \\
G + \iou{0}{G} - \iou{0}{G} & S & T & R \\ 
G & S & T & R \\ \hline
\end{array} \]
\caption{Resource permutation facilitated by agent providing outside representative money and engaging in resource trading}
\label{outside-money-exchange-iou}
\end{figure}

Note that at any given point Agent $0$ has enough $G$ in storage to instantaneously redeem the IOUs she has issued and are circulating amongst other agents; that is, the total number of $G$ is always at least as great as the number of $-\iou{0}{G}$ on her books.  As long as that is the case every outstanding IOU can effectively be treated as a certificate of ownership of the corresponding amount of $G$; Agent $0$ only \emph{possesses} (stores) the $G$, but does not \emph{own} it.

\section{Loans}

In Figure~\ref{outside-money-exchange-iou} Agent $0$ interacts with 2 other agents: Agent $1$ whom she initially transfers $\iou{0}{G}$ to in exchange for physical resource $R$ and Agent $3$ whom she eventually exchanges $R$ and $\iou{0}{G}$ with.  In other words, Agent $0$ needs to handle physical resources and engage in trades with multiple agents even though her net interest and function is to provide liquidity only.  We can remove the first need by Agent $0$ accepting an IOU from Agent $1$ instead; see Figure~\ref{outside-money-exchange-iou-collateral}.

\begin{figure}
\[ \begin{array}{|llll|}
\hline 0 & 1 & 2 & 3 \\ \hline
G & R & S & T \\
G + \iou{0}{G} - \iou{0}{G} & R + \iou{1}{R} - \iou{1}{R} & S & T \\
G + \iou{1}{R} - \iou{0}{G} & R + \iou{0}{G} - \iou{1}{R} & S & T \\
G + \iou{1}{R} - \iou{0}{G} & R + S - \iou{1}{R} & \iou{0}{G} & T \\
G + \iou{1}{R} - \iou{0}{G} & R + S - \iou{1}{R} & T & \iou{0}{G} \\
G + \iou{0}{G} - \iou{0}{G} & R + S - \iou{1}{R} & T & \iou{1}{R} \\
G & \iou{1}{R} + S - \iou{1}{R} & T & R \\
G & S & T & R \\ \hline
\end{array} \]
\caption{Resource permutation facilitated by agent providing outside representative money and trading IOUs}
\label{outside-money-exchange-iou-collateral}
\end{figure}

At every point in time all outstanding IOUs by all agents are covered by actual resources and still work as certificates of ownership.  At the end, Agent $0$ is involved in what is effectively a three-way exchange: Agent $3$ is interested in $R$, which is in possession of Agent $1$, and has money $\iou{0}{G}$ for it; $R$, however, is pledged away to Agent $0$.  So Agent $3$ acquires the right to $R$ from Agent $0$ and eventually insists on getting it delivered.  This requires a tricky three-way consensus.  

One way of breaking this three-way exchange into independent two-way exchanges is by Agent $0$ and Agent $1$ entering into a \emph{loan} agreement and, separately and independently, Agent $1$ and Agent $3$ into an ordinary goods-for-money exchange. At its core, a loan consists of (a contract to perform) two exchanges: $G$ for $\iou{k}{G}$ and then $\iou{k}{G}$ for $G$, money now for money later.  In Figure~\ref{loan-with-no-security}, Agent $0$ exchanges $G$ for $\iou{1}{G}$ with Agent $1$, and they eventually perform the reverse exchange.

\begin{figure}
\[ \begin{array}{|llll|}
\hline 0 & 1 & 2 & 3 \\ \hline
G & R & S & T \\
G & R + \iou{1}{G} - \iou{1}{G} & S & T \\
\iou{1}{G} & R + G - \iou{1}{G} & S & T \\
\iou{1}{G} & R + S - \iou{1}{G} & G & T \\ 
\iou{1}{G} & R + S - \iou{1}{G} & T & G \\ 
\iou{1}{G} & G + S - \iou{1}{G} & T & R \\ 
G & \iou{1}{G} + S - \iou{1}{G} & T & R \\ 
G & S & T & R \\ \hline
\end{array} \]
\caption{Loan of commodity money with no security}
\label{loan-with-no-security}
\end{figure}

Instead of exchanging commodity money $G$, representative money $\iou{b}{G}$ can be used, if the issuer $b$ of the promissory note is generally trusted to deliver $G$ upon demand to anybody holding $\iou{b}{G}$; see Figure~\ref{full-reserve-loan-agreement}.  

\begin{figure}
\[ \begin{array}{|llll|}
\hline 0 & 1 & 2 & 3 \\ \hline
G & R & S & T \\
G + \iou{0}{G} - \iou{0}{G} & R + \iou{1}{G} - \iou{1}{G} & S & T \\
G + \iou{1}{G} - \iou{0}{G} & R + \iou{0}{G} - \iou{1}{G} & S & T \\
G + \iou{1}{G} - \iou{0}{G} & R + S - \iou{1}{G} & \iou{0}{G} & T \\ 
G + \iou{1}{G} - \iou{0}{G} & R + S - \iou{1}{G} & T & \iou{0}{G} \\ 
G + \iou{1}{G} - \iou{0}{G} & \iou{0}{G} + S - \iou{1}{G} & T & R \\ 
G + \iou{0}{G} - \iou{0}{G} & \iou{1}{G} + S - \iou{1}{G} & T & R \\ 
G & S & T & R \\ \hline
\end{array} \]
\caption{Loan of representative money with no security}
\label{full-reserve-loan-agreement}
\end{figure}

This means a loan consists of Agents $b$ and $k$ exchanging IOUs with each other twice, with some time in between so that $k$ \emph{and others} can engage in useful exchanges in that time.  The point of the IOU exchanges is to inject liquidity into a system of trading agents by temporarily replacing a low-quality IOU $\iou{k}{G}$ by a high-quality IOU $\iou{b}{G}$ with medium-of-exchange status.

Between the two exchanges Agent $b$ holds an $\iou{k}{G}$ without $k$ actually owning $G$ and thus risks getting neither $G$ nor $\iou{b}{G}$ back.  The value and business of Agent $b$ is thus performing the interconnected tasks of providing sufficient liquidity for others to effect useful trades and taking on, aggregating and managing the risk of not getting its money back.  In this fashion she isolates other agents from having to perform their own creditworthiness checks on agents issuing IOUs.  She charges a fee (interest) for the value of that, of course, which in our examples is left out for simplicity.  And she may require a \emph{collateral}, an IOU by the borrower of a rarely traded resource $C$, as security.\footnote{Additionally or alternatively, $b$ may also require getting whatever $k$ buys with the borrowed money as collateral.}  For example, in Figure~\ref{loan-with-collateral} Agent $1$ gives $C$ as collateral to Agent $0$ during the loan.  

If Agent $1$ fails to repay $\iou{0}{G}$, Agent $0$ can insist on exchanging $C$ for $\iou{1}{G}$ and canceling (exchanging) the mutual IOUs $\iou{1}{C}$ for $\iou{0}{C}$; see Figure~\ref{loan-with-collateral-and-default}.

\begin{figure}
\[ \begin{array}{|llll|}
\hline 0 & 1 & 2 & 3 \\ \hline
G & C + R & S & T \\
G + \iou{0}{G} - \iou{0}{G} + \iou{0}{C} - \iou{0}{C} & C + R + \iou{1}{G} - \iou{1}{G} + \iou{1}{C} - \iou{1}{C} & S & T \\
G + \iou{1}{G} - \iou{0}{G} + \iou{1}{C} - \iou{0}{C} & C + R + \iou{0}{G} - \iou{1}{G} + \iou{0}{C} - \iou{1}{C} & S & T \\
G + \iou{1}{G} - \iou{0}{G} + \iou{1}{C} - \iou{0}{C} & C + R + S - \iou{1}{G} + \iou{0}{C} - \iou{1}{C} & \iou{0}{G} & T \\
G + \iou{1}{G} - \iou{0}{G} + \iou{1}{C} - \iou{0}{C} & C + R + S - \iou{1}{G} + \iou{0}{C} - \iou{1}{C} & T & \iou{0}{G} \\
G + \iou{1}{G} - \iou{0}{G} + \iou{1}{C} - \iou{0}{C} & C + \iou{0}{G} + S - \iou{1}{G} + \iou{0}{C} - \iou{1}{C} & T & R \\
G + \iou{0}{G} - \iou{0}{G} + \iou{0}{C} - \iou{0}{C} & C + \iou{1}{G} + S - \iou{1}{G} + \iou{1}{C} - \iou{1}{C} & T & R \\
G & C + S & T & R \\ \hline
\end{array} \]
\caption{Loan of representative money with collateral}
\label{loan-with-collateral}
\end{figure}

\begin{figure}
\[ \begin{array}{|llll|}
\hline 0 & 1 & 2 & 3 \\ \hline
G & C + R & S & T \\
G + \iou{0}{G} - \iou{0}{G} + \iou{0}{C} - \iou{0}{C} & C + R + \iou{1}{G} - \iou{1}{G} + \iou{1}{C} - \iou{1}{C} & S & T \\
G + \iou{1}{G} - \iou{0}{G} + \iou{1}{C} - \iou{0}{C} & C + R + \iou{0}{G} - \iou{1}{G} + \iou{0}{C} - \iou{1}{C} & S & T \\
G + \iou{1}{G} - \iou{0}{G} + \iou{1}{C} - \iou{0}{C} & C + R + S - \iou{1}{G} + \iou{0}{C} - \iou{1}{C} & \iou{0}{G} & T \\
G + \iou{1}{G} - \iou{0}{G} + \iou{1}{C} - \iou{0}{C} & C + R + S - \iou{1}{G} + \iou{0}{C} - \iou{1}{C} & T & \iou{0}{G} \\
G + C - \iou{0}{G} + \iou{0}{C} - \iou{0}{C} & \iou{1}{G} + R + S - \iou{1}{G} + \iou{1}{C} - \iou{1}{C} & T & \iou{0}{G} \\
G + C - \iou{0}{G} & R + S & T & \iou{0}{G} \\ \hline
\end{array} \]
\caption{Loan of representative money with collateral and eventual default}
\label{loan-with-collateral-and-default}
\end{figure}

\section{Inside money}

Note that $G$ is never transferred in Figures~\ref{full-reserve-loan-agreement}, \ref{loan-with-collateral} and \ref{loan-with-collateral-and-default}. If another agent ever wanted to redeem $\iou{0}{G}$, Agent $0$ could say: ``Why do so?  You are not really interested in $G$ itself for consumption or production, but just in its medium-of-exchange value.  You may as well use $\iou{0}{G}$ instead and keep $G$ safely stored with me.''  In general, if Agent $0$ owns $G$ initially and finally, every sequence of exchanges involving $G$ can be simulated by a sequence using $\iou{0}{G}$ instead of $G$.  In particular, \emph{the $G$ is never exchanged}.   
 
This gives rise to a Jedi trick idea: $G$ does not even need to exist for the purposes of exchange, only the belief in its existence!  To wit,  Figure~\ref{unsecured-loan} shows Agent $0$ issuing $\iou{0}{G}$ without actually owning $G$.  The exchanges work as before, if the other agents trust the exchange value of $\iou{0}{G}$ as much as $G$ \emph{and never ask for delivery of $G$}.  

\begin{figure}
\[ \begin{array}{|llll|}
\hline 0 & 1 & 2 & 3 \\ \hline
0 & R & S & T \\
\iou{0}{G} - \iou{0}{G} & R + \iou{1}{G} - \iou{1}{G} & S & T \\
\iou{1}{G} - \iou{0}{G} & R + \iou{0}{G} - \iou{1}{G} & S & T \\
\iou{1}{G} - \iou{0}{G} & R + S - \iou{1}{G} & \iou{0}{G} & T \\ 
\iou{1}{G} - \iou{0}{G} & R + S - \iou{1}{G} & T & \iou{0}{G} \\ 
\iou{1}{G} - \iou{0}{G} & \iou{0}{G} + S - \iou{1}{G} & T & R \\ 
\iou{0}{G} - \iou{0}{G} & \iou{1}{G} + S - \iou{1}{G} & T & R \\ 
0 & S & T & R \\ \hline
\end{array} \]
\caption{Loan without reserves}
\label{unsecured-loan}
\end{figure}

Observe that $G$ is not traded at all, only IOUs involving $G$.  
The promissory note $\iou{0}{G}$ created by the issuer before she actually has the $G$, if ever, is \emph{inside money}.\footnote{The classical distinction between outside and inside money is a question of which subset of agents is considered. The same money that is outside money in the system \emph{without} Agent 0 is inside money in the same system \emph{with} Agent 0.}

\section{Fractional reserve banking}

Issuing representative money that must be fully backed by the underlying resource at all times, as in Figure~\ref{full-reserve-loan-agreement}, is essentially \emph{full reserve banking}: all issued IOUs can instantaneously be redeemed by Agent $0$.  In particular, the underlying resources must have been produced.  
The IOU $\iou{0}{G}$ issued by Agent $0$ as part of her loan to Agent $1$ in Figure~\ref{full-reserve-loan-agreement} is an example of a  transfer that is fully covered by Agent $0$'s reserves.  If we think of Agent $0$ as a bank, it corresponds to the bank lending some of its \emph{equity} to Agent $1$.  The volume of loans it can make in this fashion is limited by its equity.

The diametric opposite to full reserve banking is if none of the IOUs are backed and thus none of them can be redeemed, as in Figure~\ref{unsecured-loan}; this is the essence of fiat money, which we will get to later.  In Figure~\ref{unsecured-loan} Agent $0$ simply ``prints'' representative money out of thin air with no reserves, loans it to Agent $1$ and gets away with it since nobody ever redeems it. 
If the volume of IOUs issued is not bounded in any fashion, that is there is a $0\%$ reserve, capital and balance sheet requirement
the bank could, in principle, print as much money as its printers will allow. 

In between full-reserve and no-reserve banking is \emph{fractional reserve banking}.
The basic idea is to have enough reserves on hand to satisfy actual redemption requests and thus bluff all IOU holders into believing that one \emph{could} satisfy any incoming redemption requests, but then count on the IOUs being mostly used as media of exchange that are eventually traded back to the issuer without ever being redeemed, thus making the sufficiency of the fractional reserves self-fulfilling.  

\subsection{Current promises backed by future promises}

A closer look at these examples reveals that the bank, Agent $0$, is not simply printing money without backing in \emph{any} resources, but it has \emph{other} resources than $G$.  In Figure~\ref{unsecured-loan}, she owns $\iou{1}{G}$, the repayment promise by Agent $1$, instead of a $G$ to cover the outstanding IOU $\iou{0}{G}$.  

There are three fundamental differences and attendant potential problems with having $\iou{1}{G}$ instead of $G$ as a reserve for an outstanding $\iou{0}{G}$.
\begin{itemize}
\item The resource $\iou{1}{G}$ is simply something else than what a holder of $\iou{0}{G}$ expects upon redemption, namely the resource $G$ itself.  The holder of a promise to get a specific refrigerator delivered on demand may not accept getting offered a freezer instead.
\item The resource $\iou{1}{G}$ is a \emph{future} payment by Agent $1$; the promise exists now, but the actual loan repayment is at some time in the future.  This form of ``reserve'' does not have to exist yet as a physical resource, but can be something that (presumably) only exists in the future.
\item The value of $\iou{1}{G}$ is uncertain; Agent $1$ may go bankrupt and default on her loan.  The future resource may never materialize.
\end{itemize}
 
Fractional reserve banking can be thought of as a form of full reserve banking with uncertainties, where the reserves are, for the most part, different resources than what the issued IOUs actually promise; the resources may exist physically only at some point in the future, if ever, and there is a risk they never will.

Relaxing full to partial reserve requirements alleviates the need for having extraordinary amounts of commodity money $G$ to realize the full potential of economic exchanges. A $10\%$ reserve requirement reduces the requirements on how many units of $G$ need to exist and be stored at any given time by a factor of $10$.  

\subsection{Seigniorage}

Fractional reserve banking gives rise to a lucrative business: \emph{Seigniorage}, the issuance (printing) of multiple IOUs for a single actual resource in reserve, where each IOU is traded at the same value as the resource itself.  
With a $10\%$ reserve requirement, IOU production has a whopping 90\% lower cost than producing the underlying resource.


\subsection{Fractional reserve banking risks}

Fractional reserve banking carries a number of risks.
 
\begin{itemize}
\item Lack of stability: A multiplicative increase of IOUs due to increased production of $G$ may devalue both $G$ and IOUs in a multiplicative fashion.  IOUs may rapidly lose value due to a bank run (see below).  A bank can go bankrupt at any point in time, making IOUs issued by that bank suddenly worthless.
\item Distortion of economic activity: The immense profitability of seignorage favors banking over other economic sectors, especially if seignorage profits are privatized, as they are in modern fractional reserve banking.
\item Bank runs: If there is a rumor, whether true or not, that a bank may not honor all IOUs it has issued, the optimal strategy for an IOU holder is to redeem their IOUs as quickly as possible, which \emph{necessarily} leads to a \emph{bank run} where \emph{all} IOU holders seek to redeem their IOUs.  This is, by definition of fractional reserve banking, not possible.  In this case some IOU holders will get their redemption request satisfied, some not.  In particular, IOUs are not arbitrarily interchangeable, not living up to the fungibility requirement of money.  It depends on how fast their owners are at redeeming them.   If, at some point, it is decided to reduce the value of an IOU to the fraction actually covered by existing reserves, as in a bank default, the IOUs issued by that bank lose exchange value, which negates their presumed stability-of-value property. 
Such reduction indirectly converts seigniorage profits made earlier and transferred (as dividends) to the owners to losses for those holding the IOUs at the time of the bank run.\footnote{In practice a bank experiencing a run has other assets, as we noted, which affects the effective value of an IOU after a bank's bankruptcy.} 
Furthermore, as the IOUs issued by the particular bank suddenly become worthless and IOUs issued by other banks retain their exchange value, this negates the value of bank-issued IOUs as \emph{single} unit of account that is independent of their issuer.  

\end{itemize}

These are not just hypothetical risks.  The world-wide depression of 1929 started with a bank run in the US where depositors demanded their deposits paid out in US Federal Reserve issued cash. The banks employed fractional reserve banking, so they did not have enough cash reserves covering the customer deposits.  
They tried to borrow cash from The Federal Reserve, which issues the cash.  Since cash had a full reserve requirement in terms of gold, the Federal Reserve did not have nor procure enough gold to print enough cash to make the loans to the banks.  The banks defaulted; other banks who had given credit to them (had IOUs from them), saw those IOUs become worthless, defaulted themselves and so on.   

\section{Fiat money}

An agent $k$ can create $\iou{k}{G}$ (and its dual $-\iou{k}{G}$) out of thin air instantaneously, in the desired amounts and on demand.  In both full reserve and fractional reserve banking redemption of an IOU $\iou{k}{G}$, that is exchanging it for the resource $G$, requires that $k$ has such $G$ on hand.  What if we simply remove the redemption part of an IOU? 
Then nobody can show up,  $\iou{k}{G}$ in hand, and demand delivery of $G$; the IOU can only be used as medium of exchange, \emph{not} for redeeming it (exchanging it for delivery of $G$).   
As we have seen in Figure~\ref{unsecured-loan} the IOUs $\iou{k}{G}$ are sufficient for facilitating multiway exchanges, without $G$ being on hand or even existing.  $G$ could be Rai stones, unicorns, dilithium on Planet Remus or anything else that is entirely made up. 
What makes an unredeemable IOU $\iou{k}{G}$ be perceived as money is the collective trust in $\iou{k}{G}$---and thus Agent $k$ as its issuer---having and retaining a \emph{stable} exchange value over time and across space when trading it against ordinary economic resources such as $R, S, T$ in Figure~\ref{unsecured-loan} that are used for consumption and production.  If and only if this succeeds, this is \emph{fiat money}\footnote{From ``fiat'', Latin, for ``Let it be done.''}: it is made up and it has store-of-value, stable-value and unit-of-account properties of money.  The key question, of course, is: how can a non-redeemable IOU, essentially a piece of paper that says ``Worth 5 unicorns. Signed, Alice'', be trusted to be a store of value and serve as a medium of exchange?  

\section{Central bank money}

A \emph{central bank} is a designated agent $0$ that is \emph{exclusively} able and empowered by a sovereign state to produce nonredeemable IOUs $\iou{0}{M}$ of some  made-up resource $M$, called a \emph{currency}, out of thin air; that is, it can split $0$ into any number of units of $\iou{0}{M}$ and its dual liability $-\iou{0}{M}$ at any time and subsequently transfer $\iou{0}{M}$ (but not $-\iou{0}{M}$) to any other agent. 

An example of Agent $0$ is Nationalbanken, the central bank of Denmark, where $M$ is $\DKK$.  Let us call $\DKK$ the Fabled Danish Krone; it is entirely made up. The central bank can instantaneously print a promissory note $\iou{0}{\DKK}$ saying ``The bearer of this note can redeem it at any time for 1 Fabled Danish Krone at Platform $9 \frac{3}{4}$, Kings Cross Station.  Signed, Nationalbanken.'' and simultaneously print a note ``Nationalbanken owes 1 Fabled Danish Krone to whoever redeems such note'' and put it in its own ledger.  It can then give or loan the promissory note to another agent $k$, who in turn can use it as medium of exchange to pay other agents for resources she needs.  In case of a loan, $k$ eventually needs to trade resources she has produced to get back a note issued by the central bank and return it at the end of the loan.  Importantly, such note need not be the \emph{same} note; any note (or notes) with the same (aggregate) amount can be substituted for the original note; that is, notes are \emph{fungible}: any set of notes can be replaced by any other set of notes with the same aggregate amount of Fabled Danish Kroner.  Otherwise, each note would ultimately constitute its own currency with potentially different value to different agents, diminishing its role as liquid medium of exchange.  

Mathematically, starting with $0$, the sum total of the values of IOUs in existence at any given point in time, including the liabilities of the form $-\iou{0}{M}$, remains $0$.  This is because both transforming $0$ into $\iou{0}{M} + (-\iou{0}{M})$ and transferring $\iou{0}{M}$ from one agent to another keeps their sum total across all agents invariant, and IOUs can be neither lost nor duplicated.\footnote{We can model lost money as being owned by a hypothetical ``sink'' agent that only receives money without ever paying anybody, and successful forging as unauthorized issuance in the central bank's name.} Since only the central bank $0$ can own liabilities $-\iou{0}{M}$, this means that it is the only agent 
with a negative balance, which necessarily equals the sum total of IOUs held by all other agents combined.   The key job of the central bank is to observe \emph{prices}, the amount of $\iou{0}{M}$ paid in exchanges with goods and services and, in some fashion, add or extract amounts of $\iou{0}{M}$ circulating amongst the other agents to keep those prices stable over time.  (The ``in some fashion'' is the tricky part, of course.)

To position $\iou{0}{M}$ as a dominant medium of exchange, an important agent, the \emph{government}, can force other agents to make payments (think taxes) in $\iou{0}{M}$ as well as insisting on making payments itself in $\iou{0}{M}$. This creates an immediate demand for $\iou{0}{M}$ and self-reinforcing trust in its function as a canonical medium of exchange.  Additionally, legislation may be put in place that forces all agents in a country to accept payment in $\iou{0}{M}$, ensuring that an agent can pay with $\iou{0}{M}$ down the line when accepting $\iou{0}{M}$ in a trade. Additionally, disallowing payment in competing money-like resources eliminates potential competitors that could fragment liquidity and undermine a government's and central bank's fiscal and monetary policy power.

\section{Private bank money}

Any agent $b$ can in principle issue transferable promissory notes and thus produce a personal currency.   For example, Bob can issue a note that says that the bearer can redeem it for a central bank note like the above; that is, the note represents $\iou{b}{\iou{0}{\DKK}}$.\footnote{``The bearer of this note can redeem it at any time at my office for a note issued by Nationalbanken that says that its bearer can redeem it for 1 Fabled Danish Krone at Platform $9 \frac{3}{4}$, Kings Cross Station.  Signed, Bob.''}  The typical form is a check to ``Bearer'' (or ``Cash'') to be paid in cash (central bank money).  Such a promissory note is \emph{not} fiat money; indeed its prime purpose is being redeemed for the central bank note, which does constitute fiat money as it cannot be redeemed.  Being transferable, Bob's note \emph{may} serve as medium of exchange -- it is Bob's Danish Krone (BDKK) -- but it carries a serious risk of redemption failure since Bob may issue many more such notes than he can assuredly redeem for central bank money.  For example, Bob can issue new BDKK, transfer them to his friends in exchange for promises to get them back later (loans), have his friends redeem BDKK for all the central bank DKK Bob actually has, and then declare bankruptcy, leaving the remaining holders of BDKK without possibility of redeeming them for central bank DKK, especially if  Bob's friends in the mean time have lost the central bank DKK they got in a casino (they say).

Private bank money in the form of \emph{bank deposits} is like Bob money: promissory notes by a \emph{regulated} agent that are redeemable for central bank money (cash) upon demand.
Regulation and oversight of a bank are there to avoid that a bank's owners run from their obligations by declaring bankruptcy or print too many IOUs that they cannot redeem. 

\section{Central bank money versus private bank money versus private IOUs}

The differences between $M$, $\iou{c}{M}$, $\iou{b}{M}$ and $\iou{k}{M}$ where $c$ is the central bank, $b (\neq c)$ is a bank, and $k$ is neither central bank nor ordinary bank are subtle, but important:
\begin{itemize}
\item 
$M$ is \emph{fiat currency}, a mythical resource that can only be manufactured by the central bank $c$; it can do so near-instantaneously, at near-zero cost and in arbitrary amounts.  
\item $\iou{c}{M}$ is fungible \emph{central bank currency}; an unnumbered IOU in some amount issued by the central bank to deliver $M$ to its holder on demand, but with the  extra clause that it \emph{cannot} be redeemed.  An amount of $\iou{c}{M}$ corresponds to any set of unnumbered bills or coins whose sum adds to the given amount.  Since $\iou{c}{M}$ cannot be redeemed it is indistinguishable in function from $M$.  Indeed, we will identify $M$ and $\iou{c}{M}$:
$$M = \iou{c}{M}$$
Mathematically, we can think of one $M$ as the infinite IOU sequence $\iou{c}{\iou{c}{\ldots}}$.  It is an IOU that, when presented to the central bank for redemption, is simply returned: ``Here is what you are owed.'' This model explains why $\iou{c}{M}$ is simultaneously a claim against the central bank and an asset: it can be redeemed, but only for itself.\footnote{Calling it a claim is somewhat misleading, almost facetious since one cannot really get anything else but the claim back.  Redeeming a claim against a shoemaker to get a pair shoes results in a pair of shoes being delivered. The shoes can then be used for walking, which the claim itself is not suitable for.  The claim and the shoes are observably different.  Trying to redeeming a 100 DKK note at Nationalbanken is a different matter.  At best it may result in being exchanged for two 50 DKK notes. Whatever is possible with those can also be done with the 100 DKK note; they are observably equivalent.}  
Instances are \emph{account balances}, whether held in a
centralized or decentralized database such as the currency reserve balances in a central bank database system or the Ether balances associated with each address in Ethereum, where ``Ethereum'' is the (decentralized implementation of the) non-sovereign ``central bank'' issuing the magical currency Ether and ``Bitcoin'' is the (decentralized implementation of the) non-sovereign central bank issuing the magical currency Bitcoin.  
\item $M_n$ is a \emph{currency note}, a numbered (serialized) central bank note in some particular amount. The serial number $n$ can be used to track its use; in particular, it is distinguishable from other notes with the same amount, but different number.  Examples are numbered (physical) cash notes issued by a central bank, (digital) \emph{unspent transaction outputs (UTxO)} in Bitcoin-style blockchain and distributed ledger systems, and nonfungible tokens in Ethereum-like blockchain systems. 
\item $\iou{b}{M} (= \iou{b}{\iou{c}{M}})$ is \emph{bank money}, an IOU issued by bank $b$ that can be redeemed for $M$.  It is a legal requirement that $b$ must fully honor the IOU: it must deliver $M$ in exchange for receiving $\iou{b}{M}$ at a 1:1 bank money for central bank currency exchange rate.  Note, though, that $b$ may fail to honor the redemption request; it may simply default on it.   Furthermore, $\iou{b}{M}$ and $\iou{b'}{M}$ for banks $b, b'$ are required to be exchanged at a 1:1 exchange rate, which, modulo default by $b$ or $b'$, is a corollary of $\iou{b}{M}$ and $M$ being one-to-one exchangeable.\footnote{Bank $b$ may, however, want to refuse to enter into a contract with a customer where it issues an $\iou{b}{M}$ in exchange for $M$, that is let customers deposit physical cash.  This might be the case where, as is currently the case, banks receive less interest from the central bank than it gives its customers.} 
\item $\iou{k}{M}$ is a promise by $k$ to redeem it for $M$. Depending on who $k$ is it may be worth anything between $M$ and nothing.
\end{itemize}
These are all transferable \emph{bearer instruments}.

Note that an ordinary private bank account is \emph{not} a specific amount of central bank currency owned by its account owner and merely stored in the bank.  The account owner owns an IOU issued by the bank, not the central bank, in the amount of its current balance.  The difference is that the account owner does not own the central bank currency itself, but only the \emph{promise} to get the central bank currency from the bank \emph{on demand} -- the bank may not (and typically will not) actually \emph{have} enough central bank currency on hand to redeem the IOU on demand.  

It is instructive to illustrate the difference between bank money and (central bank issued) physical cash.
The cash in a bank box is central bank currency owned by the customer who put it in the bank box; the cash is \emph{not} owned by the bank.  The bank merely provides the storage facility.  If the cash disappears from the bank box without permission by the customer, somebody has \emph{stolen} it, which is a criminal act.  A standard bank account, on the other hand, is a \emph{call loan} made to the bank: The customer transfers ownership of the central bank currency to the bank in exchange for a \emph{promise} to get it returned on demand.  If the customer makes such a demand and the bank does not comply with it, this is ``only'' breach of contract, not a criminal act.  If a bank goes bankrupt and it is found that the cash contents of its bank boxes have disappeared because the bank used it for other purposes, then this constitutes theft.  A bank account balance that cannot be honored, on the other hand, is failure to satisfy a contract, which is not theft.  Account owners provide loans to the bank; thus, in principle, they should conduct a thorough creditworthiness involving \emph{everything} the bank engages in to assess their chances of getting their loans back. 

Banks offer bank boxes in their vaults for securely storing valuables, including \emph{physical} cash.  A natural question is: Why not offer secure storage of central bank \emph{digital} currency?

\chapter{Keeping track of money}
\label{keeping-track}

\chapterauthor{Fritz Henglein}

In Chapter~\ref{exchange-theory} we started out with a conceptual ownership state: agents $0, 1, 2$ own resources $R, S, T$, respectively, and engage in exchanges eventually resulting in them owning $T, R, S$, respectively. But how is this ownership state implemented in practice?  Most importantly, how is it ensured that transfers of resources are really transfers: that the resource is subsequently not only owned or possessed by the recipient, but is also no longer owned or possessed by the sender?  

This is a particularly touchy issue for fiat money: Fiat money is entirely made up.  Its greatest achievement, being extremely cheap to produce is also its greatest achilles heel: how to prevent it from being duplicated (sending the money, but keeping a copy -- that is forging it) or, maybe of slightly lesser concern, being lost?  

More drastically, with entirely digital fiat money, there is not even a basement to store physical bills and coins in to cross-check how much money there really \emph{is}. Where is it then? In a spreadsheet, database, blockchain system, smart card chip?  How can one be sure that somebody updating a number in a spreadsheet in some bank doesn't create money that didn't exist before and just gives it to a good friend?  

In this chapter we discuss a conceptual framework for representing an ownership state and how money transfers are made in it.  It unifies the notions of token-based and account-based money, which are often considered distinct, and suggests that a key distinction is whether transfers are more or less confidential (privacy-preserving), which is a property of the \emph{transfers}, not of the money being transferred.

\section{Ownership states and transfers}
\label{ownership-states}

A resource exchange, the fundamental transaction in the exchange theory of Chapter~\ref{exchange-theory}, consists of 
two \emph{resource transfers}: one from A to B, the other from B to A. A \emph{resource ownership state} is represented by a \emph{map} from agents to resources.  This map is \emph{conceptual} in nature; it represents a \emph{mathematical} function that is not necessarily known (observable) in its entirety by any single agent.  Each agent has a \emph{view}---partial information---of this map; the stipulated mathematical existence of the map expresses that the views are consistent with each other, not that there is any agent who knows the entire function.
In our explanations the \emph{reader} has been put into the role of an almighty Olympic observer who can ``see'' the entire ownership state; this must not be confused with any real \emph{agent} being able to do the same.  In particular, this \emph{does not} even imply that the map necessarily exists as a table, database or other data structure in any one computer or network of computers. 

\section{Ownership: Control and balance}

It is useful to introduce an intermediate layer between agents and the resources they own: agents \emph{control} \emph{resource identifiers}, unique identifiers that are mapped to a \emph{balance} of the (economic) resources they represent at a particular point in time.  If, at a particular point in time, agent $a$ \emph{controls} resource identifier $p$ and $p$ points to resource $r$ we say that $a$ owns $r$ at that time, and the set of all ownership relations of this kind constitutes the (global) ownership state at that time.  In this fashion the functional composition of control and balance maps defines the resource ownership state.  This provides a useful conceptual as well as mathematical and computer systems-oriented framework for both physical and digital as well as token-based and account-based forms of money.

\subsubsection{Physical cash}

\begin{figure}
\begin{center}
\includegraphics[width=0.9\linewidth]{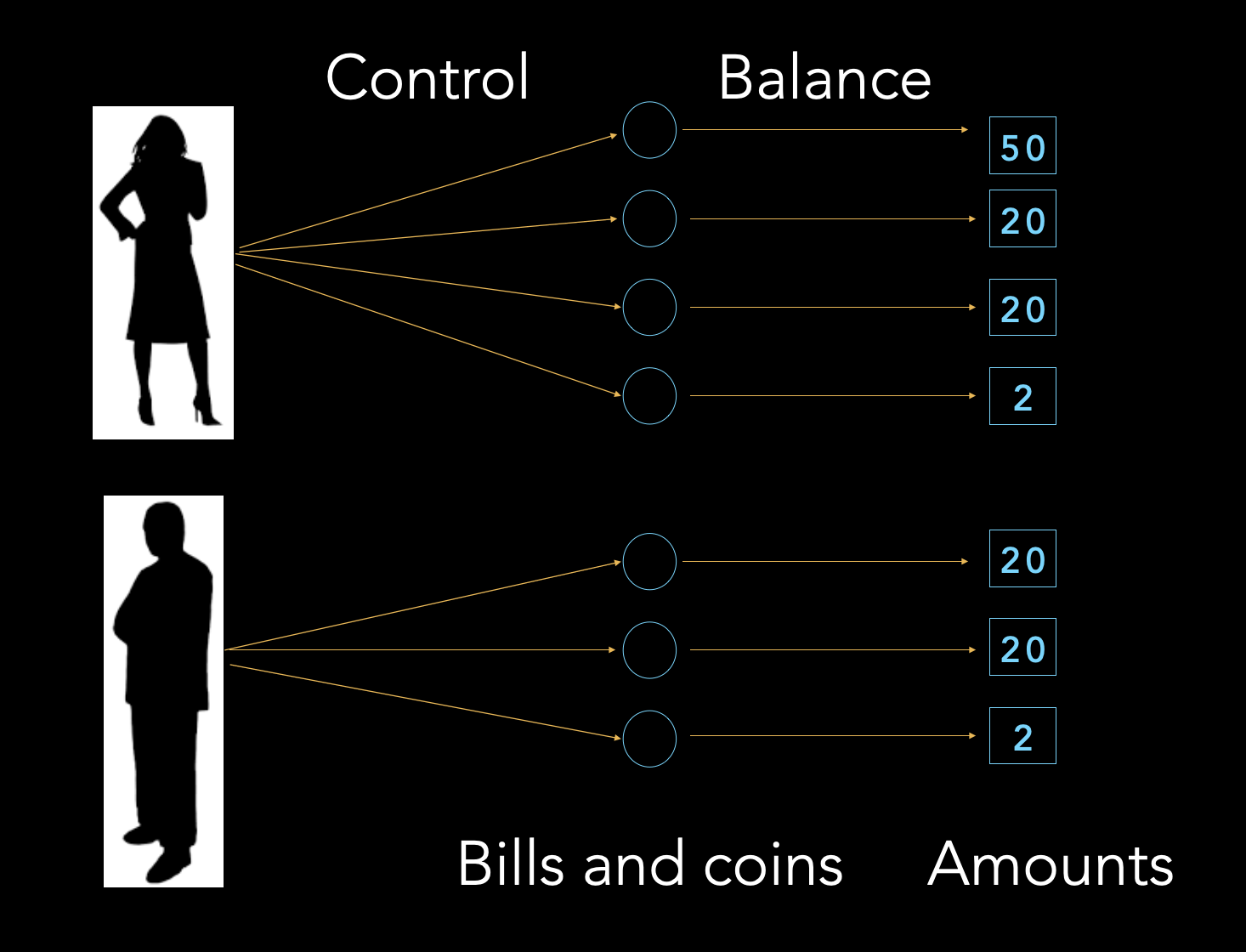}
\end{center}
\caption{Control and balance maps for physical cash}
\label{bills-and-coins}
\end{figure}

For physical cash, the resource identifiers are physical coins and bills, which are mapped to their money value, which is what is written on them.  Agents exercise control over the resource identifiers by controlling access to their location in the real world.  For example, in Figure~\ref{bills-and-coins}, Alice controls 4 distinct bills and coins by knowing where they are and being able to effectively handing them over to anybody else she and only she decides to give them to. They have balances of DKK 50, 20, 20 and 2, respectively.  The control map is completely decentralized and private; it is not stored in any single place or system: there is no single agent who knows which agent controls which resource pointers.  The balance map is also quite private: the central bank (only) knows which resource identifiers are circulating and how much each is worth; other agents don't know that beyond possibly remembering/storing which notes they happen to have owned themselves at some of time, for example by storing the serial numbers of each banknote that passes through their hands.  Note that the control and balance maps exist conceptually, but are not stored in any one place or computer system.

\subsubsection{Bank money}

\begin{figure}
\begin{center}
\includegraphics[width=0.9\linewidth]{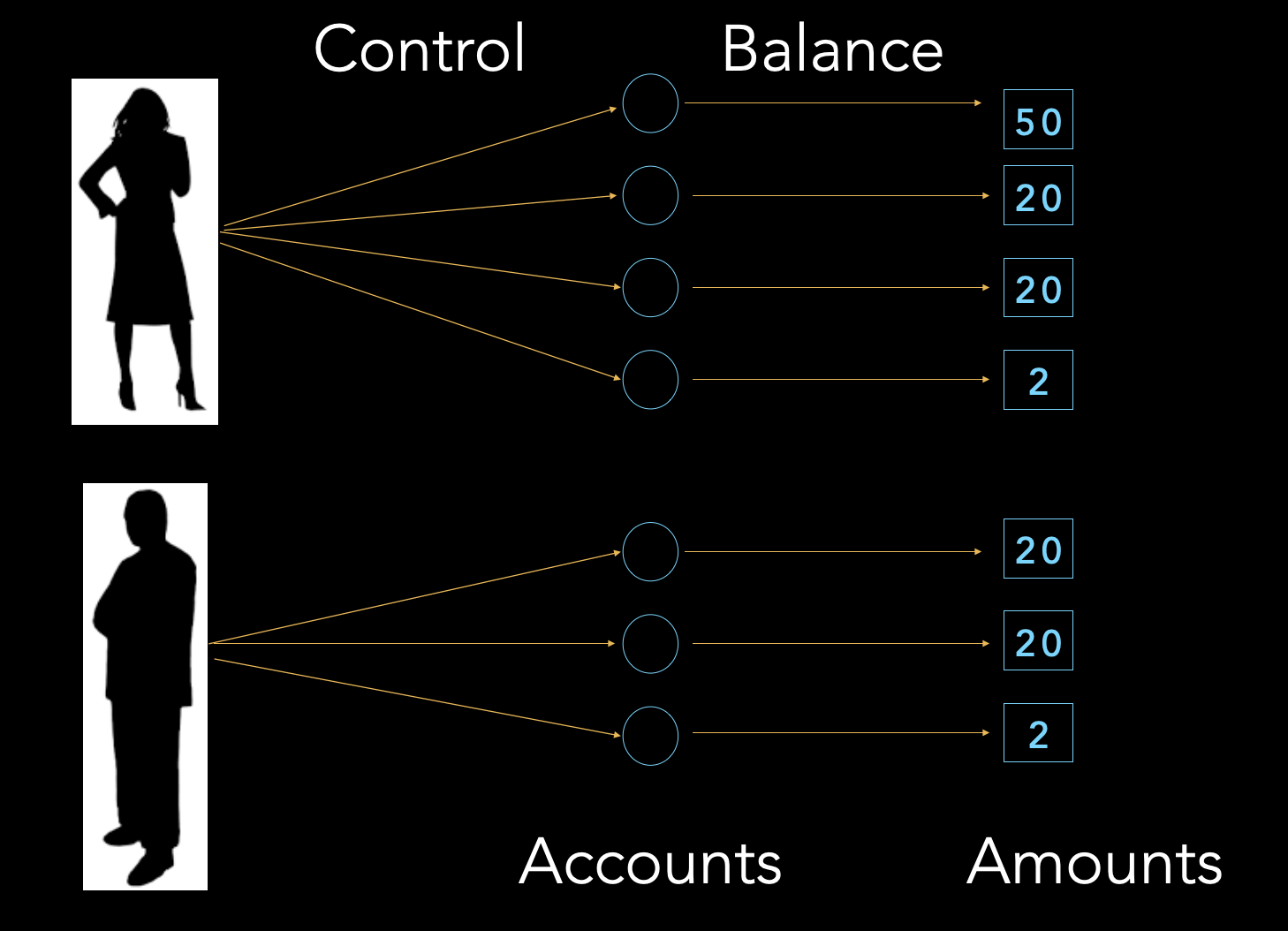}
\end{center}
\caption{Control and balance maps for bank accounts}
\label{accounts}
\end{figure}

For bank money, the resource identifiers are account numbers, which are mapped to their balances. The account numbers are partitioned into subsets, where each is stored and managed by a separate bank.   Agents exercise control over the accounts by authorizing the bank managing them to transfer an amount to another account.  For example, in Figure~\ref{accounts}, Alice controls (has) 4 bank accounts (they could be in different banks or in the same bank), with account balances of DKK 50, 20, 20 and 2, respectively.  The control map is not particularly private.  Nowadays, a bank is required to know which accounts are controlled by which person or company, and it can be compelled to disclose this information to others, notably law enforcement. A bank does not nor need not know who controls the bank accounts in other banks.  

\subsubsection{Cryptocurrency}

\begin{figure}
\begin{center}
\includegraphics[width=0.9\linewidth]{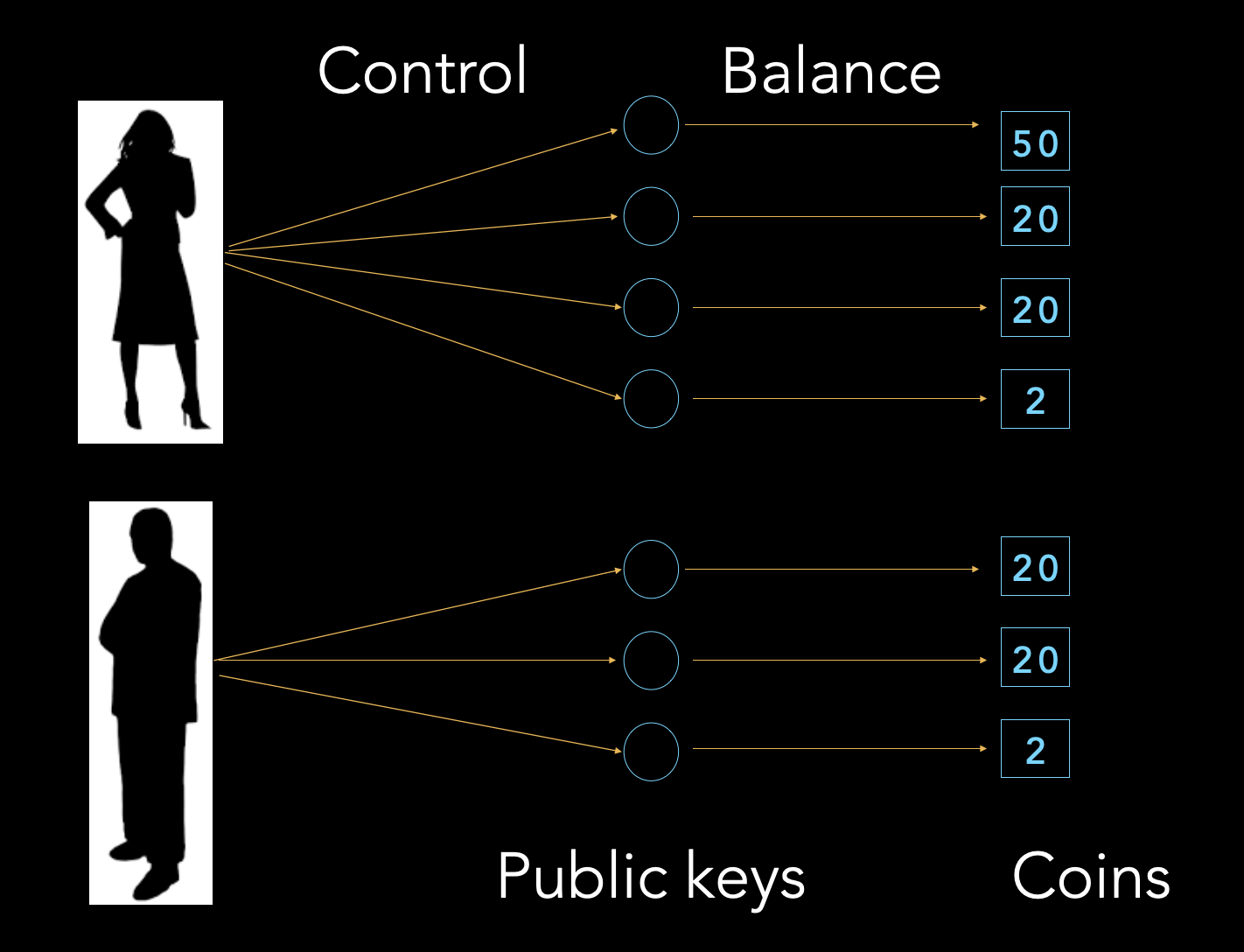}
\end{center}
\caption{Control and balance maps for cryptocurrencies}
\label{public-keys}
\end{figure}

For cryptocurrency (blockchain-hosted currency), the resource identifiers are addresses corresponding to public keys, which are mapped to their balances (Ethereum) or sets of UTxOs, where each UTxO has a fixed balance (Bitcoin).  Addresses are equivalent to account numbers in a bank.
Agents exercise control over addresses by a digital signature: providing proof of knowledge of a secret private key corresponding to the public key, without effectively revealing the private key itself or anything else about it.  The control map is
private, as for physical cash.  The balance map is completely public, however: every agent anywhere has access to it in its entirety at any time. For example, in Figure~\ref{public-keys}, Alice has 4 private keys, each corresponding to a separate public key and thus to a unique address. The balances of the 4 addresses are 50, 20, 20 and 2 Ether, respectively (Ethereum); or one UTxO each that asserts ownership (``pay to pub key hash'') of 50, 20, 20 and 2 BTC, respectively (Bitcoin). 

Functionally, Bitcoin and Ethereum can each be thought of as a bank that offers numbered accounts, where \emph{nobody} at the bank conducts a KYC check and no documentation as to the origin of the funds is required, but all account transfers, that is updates to the balance map, are posted publicly, in contrast to banks, which are expected to do the opposite: keep transfers and balances private, with the notable exception that bank employees and systems have access to them.

In contrast to Bitcoin, Ether and similar blockchain-based cryptocurrencies with public transaction ledgers, some blockchain systems additionally seek to achieve complete transaction privacy of cryptocurrency transactions.  This corresponds to a virtual bank that offers numbered accounts, has no idea about who controls them, and validates and performs encrypted money transfers in a secure enclave of its computer system such that nobody, not even at the bank, knows or can deduce which accounts and how much money is involved in a transaction or even what the balances of its accounts are.  The aforementioned computer system is virtual in blockchain systems: it is a decentralized peer-to-peer network implementing a replicated state machine that collectively acts as if it where a 1960s mainframe computer that executes single-threaded programs only, in batch mode, additionally employing various cryptographic techniques such as zero-knowledge proofs to keep information private even inside the network.\footnote{A zero-knowledge proof of a statement such as ``No money has been duplicated or lost in the transactions so far'' is a proof that is convincing to a rational -- and mathematically inclined -- observer without disclosing any additional, effectively usable information to the observer but the veracity of the statement.}

If a bank with no knowledge about their account owners' identities, performing transfers blindly and retaining no information about them sounds outrageously in conflict with banks collecting and storing extraordinary amounts of private and confidential information about individuals and companies to satisfy KYC, AML, CTF and other legislation, it is worth pointing out that Physical Cash, the (real-world) locations of physical cash around wherever it is, can be considered to be such a zero-knowledge bank.  And it is -- still -- legal in all known jurisdictions.  With physical cash there is no real-world bank, a private company with employees and computer systems under its direction, that is required and entitled to get detailed information from the account owners (cash owners), it is the account owners themselves who are responsible for providing information about their identity, storing relevant transactions and providing additional documentation in evidence of origin of funds and compliance with legislation to surrender it directly, without detour through a private third party, to law enforcement agencies on a concrete need-to-know basis. 

\section{Transfers}

The effect of a \emph{resource transfer} is to update the resource ownership state such that the ownership has changed, but the sum total of resources remains the same.  

\begin{figure}
\begin{center}
\includegraphics[width=0.9\linewidth]{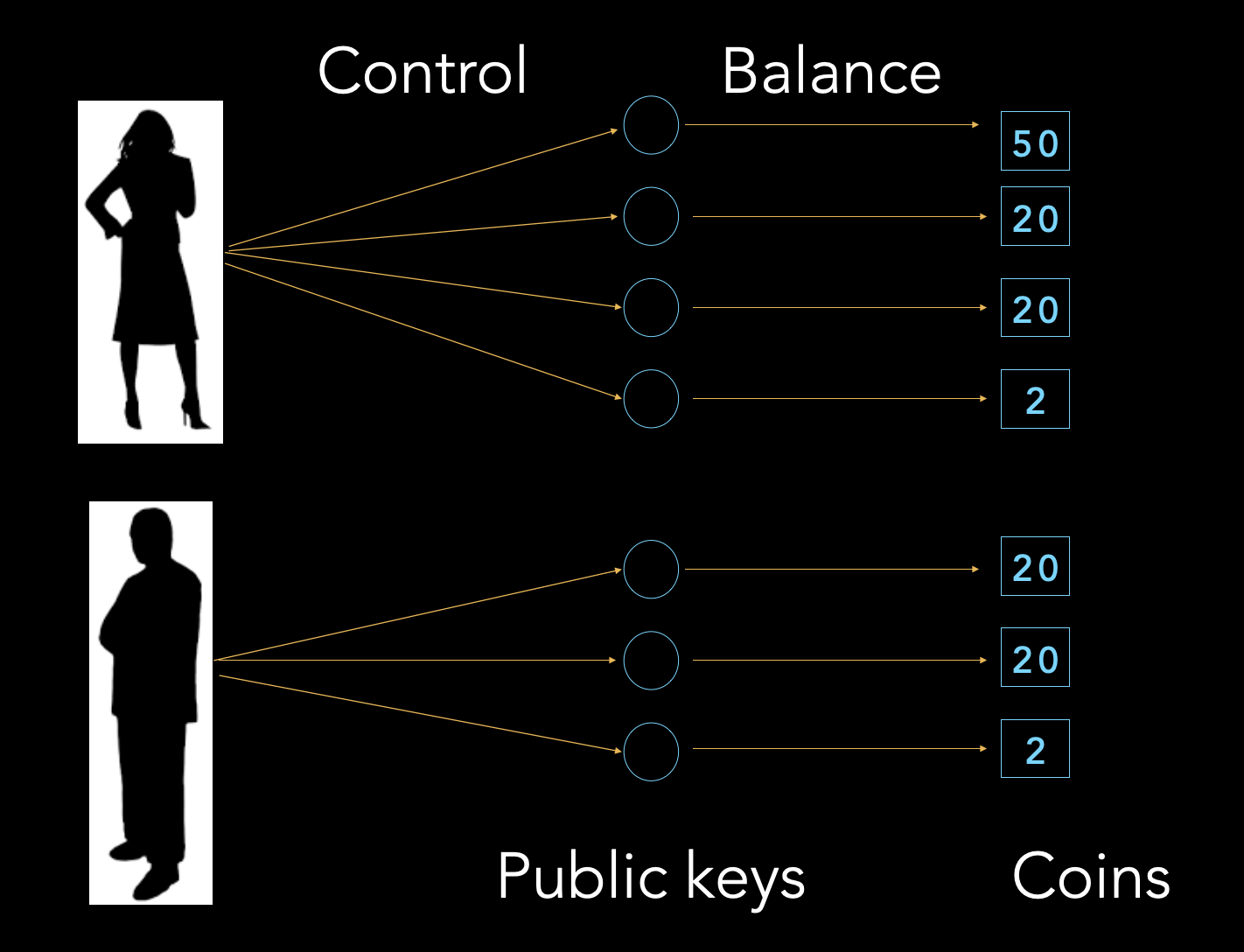}
\end{center}
\caption{Resource transfer by control or balance transfer}
\label{transfer}
\end{figure}

Having decomposed ownership into control and balance, there are two fundamental ways of achieving a transfer. 
\begin{itemize}
\item Control-based: transfer of control.  
\item Balance-based: transfer of balance.
\end{itemize}
In Figure~\ref{transfer}, Bob transfers 20 units of a currency to Alice in one of two ways. In the account-based transfer, only the balance map is updated: balance of one of the resource identifiers he controls is decreased by 20 and the balance of one of Alice's resource identifiers is increased by 20.  In the control-based transfer, only the control map is updated: Bob transfers control of one of his resource identifiers, which has a balance of 20, to Alice. 

If only control transfers are possible, then the balance of a resource pointer is fixed.  Such a resource identifier can only be created with a fixed balance and eventually retired in its entirety; it is a \emph{token}.

\section{Token-based versus account-based money}
\label{token-versus-account}

Money and other resource (asset) transfers can happen control-based or balance-based; be maximally public (all agents can see them), maximally private (only the transfer agents can see them) or 3d-party-visible (a designated agent such as the bank maintaining an account system can see them); and occur in a system with centralized or decentralized governance.  

Specific combinations are often taken as the point of departure in discussions on \emph{currency}: token-based money (physical cash) and account-based money (bank money stored in a database system); see Table~\ref{money-types}.  They can be freely combined, however: Asset transfer granularity/fungibility, privacy level and governance/distribution can in principle be designed and implemented and thus tuned to whichever requirements one has in mind.  Discussions on central bank digital currency (CBDC) need not and should not be limited by restricting assumptions about CBDC hosted on a decentralized system necessarily requiring a UTxO-model, by an account-based system necessarily having centralized governance, awarding helicopter money or paying interest requiring a database system, etc.  

For example, popular cryptocurrencies have already added some additional combinations: Unspent transaction outputs (UTxO); accounts with public balance-based transfers;  fungible (balance-based) and nonfungible (control-based) publicly transferable tokens. See Table~\ref{money-types} with examples. 

The relation between account-based and token-based money can be made precise by algebraic resource accounting \citep{torresgarcia2020}.  

\begin{table}
\begin{center}
\begin{tabular}{|l|cc|ccc|cc|} \hline
Asset transfer type  & \multicolumn{2}{c|}{Transfer} & \multicolumn{3}{c|}{Privacy} & \multicolumn{2}{c|}{Governance} \\ 
      & control & balance & publ. & 3d party & priv. & centr. & decentr.\\ \hline
Account-based  & & \checkmark & & \checkmark & & \checkmark & \\
Token-based & \checkmark & & & & \checkmark & & \checkmark \\ \hline
UTxO (Bitcoin) & \checkmark & & \checkmark & & & & \checkmark \\
Account balance (Ether) & & \checkmark & \checkmark & & & & \checkmark \\ 
Fungible tokens (ERC-20) & & \checkmark & \checkmark & & & & \checkmark \\
Nonfung.~tokens (ERC-721) & \checkmark & & \checkmark & & & & \checkmark \\ \hline
Digital cash & & \checkmark & & & \checkmark & & \checkmark \\ \hline
\end{tabular}
\end{center}
\caption{Asset transfer types}
\label{money-types}
\end{table}

\chapter{Digital cash}
\label{digital-cash}

\chapterauthor{Fritz Henglein}

In Section  \ref{token-versus-account} we discussed the most common forms of combining control and balance as a framework for describing how money is stored and transferred: token-based and account-based money. 

Intuitively, account-based money is more \emph{fungible}.  It is like token-based money with the resource identifier removed. It is analogous to removing the serial numbers from banknotes and going one step further: treating any set of banknotes as completely indistinguishable from a single unnumbered banknote that carries the sum of their values.  It is only the sum that can be observed. There is no way of distinguishing a ``collection'' of money from any other collection with the same amount: the collection immediately melts into its sum, the balance.  This maximal level of fungibility---complete indistinguishability---is the essence of account-based money, \emph{money per se}: money is money, \$50 is \$50, no matter what the history of money transactions is or into how many units (bills, coins, tokens) the amount is partitioned.  In particular, money should not come with a history of transactions that is inextricably attached to it.  Such histories make differences between money with the same face value observable and thus may give them \emph{different} exchange value.  A bitcoin used in a ransomware attack is not worth as much as a newly minted bitcoin, compromising its unit-of-account and medium-of-exchange properties.\footnote{A bitcoin is effectively a nonfungible token.  There are multiple reasons why Bitcoin can be considered money-like, but not money: it is expensive and slow to produce due to mining costs and timing (it is essentially a non-consumable commodity with high production costs), in contrast to cheaply produced fiat money; it is not a stable store of value since it has a built-in deflationary issuance policy, which is furthermore not tied to economic production for stability; it is not a unit of account and is not one-for-one interchangeable, since bitcoins have observable histories, which impacts their exchange value.}

Account-based money has practical advantages over token-based money. Changing a set of tokens (or UTxOs) into another one with the same total amount is automatic and built-in in account-based money; it does not require a transaction, as in token-based money.  A large set of transfers can be replaced by another set with the same effect, as in \emph{netting} when a set of transfers amongst two or more parties are consolidated into net flows between them.  Nobody can complain about doing that with fungible money since all sets of transfers with the same net effect are in principle \emph{identical}: there are no tokens with observable and trackable identities.\footnote{Access to a transaction history still reveals a lot, of course, even if the money \emph{per se} is fully fungible.  This is analogous to a cash coin, by itself, being indistinguishable from other cash coins and providing no information where it came from.  A transaction history contains the fungible money involved, but fungible money by itself does not point back to transactions.  In particular, fungible money exists by itself, without any transactions attached to or accessible from it.}   In contrast, netting UTxO- or NFT-based payments may lead to complaints:  Receiving a bitcoin that has previously been involved in a ransomware attack instead of a ``clean'' one with the same amount may have different value (and consequences) for its recipient.

\section{Digital cash criteria}
\label{digital-cash-criteria}

We stipulate that ideal \emph{digital cash} should be 
\begin{itemize}
\item sovereign fiat money (not private bank money), that is not an IOU that can become worthless because the issuer refuses to or cannot redeem it;
\item maximally fungible, that is account-based money; 
\item transferred with maximal privacy, without transaction histories attached; 
\item be under its owner's direct control, without requiring cooperation by a designated third party (bank or e-money account provider); and 
\item have decentralized governance, without \emph{a priori} designated agents that have privileged access to balances and control of transfers.
\end{itemize}
See the bottom line of Table~\ref{money-types}. 

Ideal digital cash emulates physical cash, with two notable differences: It is account-based (not partitioned into tokens, like coins and bills), and it is \emph{in principle} transferable in arbitrary amounts near-instantaneously and over arbitrary long distances. 
Does this mean that a particular digital cash design \emph{must} admit these properties? Of course not.  Being freed from the \emph{inherent} limitations of physical cash is to make \emph{different} designs of digital cash with \emph{particular restrictions} possible.  Different designs  may be mutually exclusive and may have different objectives, depending on the jurisdiction\footnote{or anarcho-libertarian community}. 

For example, in one digital cash design all transactions might be public (as in most blockchain systems); in another transfers up to some small amount could be completely private, large transfers disclosed or disclosable to a regulatory body, and transfer settlement time could depend on the size of the transfers (the bigger the amount, the slower settlement); in yet another, transfers could be restricted to legal entities registered in a certain jurisdiction; etc.  

These are (money) policy and (system) governance questions that should ideally be kept separate from (technical) distributed systems design and implementation to ensure that existing systems and their proponents do not unduly influence policy and governance \emph{desiderata}. Which kinds of digital cash do we, say in Denmark, want, is likely quite different from, say, Ukraine. How private, how limited, which amounts, how much slowed down, monitored by whom under which legal authority and under which circumstances, with or without support for helicopter money, with or without support for dynamic amount adjustments (interest)?  How are they governed -- who decides what about the properties of the particular digital cash design, and who has access to which information under which circumstances?  And only then: How do you build a (distributed) system that delivers this within the limits of what is theoretically and practically possible in distributed computing, cryptography, hardware design, verified programming?  

Depending on particular desiderata, the digital cash could be hosted on and operated by a secure distributed system with privileged/centralized IT-governance, e.g.~by a trusted cloud system provider; a decentralized system whose nodes are operated by an open or closed association (DLT, permissioned blockchain); a decentralized system with no authentication (permissionless blockchain); or combinations of those.  

Ideally, distributed systems design and implementation should depend on and be driven by policy and governance desiderata, not conversely. The design space of distributed systems is far from exhausted by existing centralized RDBMS- and decentralized peer-to-peer, blockchain- or DLT-based systems.

\section{Digital cash and banks} 

We have already noted that private banks are put into the somewhat peculiar---privileged and precarious---position of requiring their customers to provide confidential information that is relevant for investigating, ensuring and enforcing that the customers abide by legislation: paying their taxes, not laundering money, not funding criminal or terrorist operations.  These are the \emph{customers'} obligations vis a vis \emph{government} institutions put in charge of law enforcement.  Why are banks, as private enterprises that are primarily responsible to their shareholders, in the middle of this?  

The disassociation of digital cash from bank money and bank transactions makes it possible to deconstruct bank services.  A bank can still act as appointed custodian for customers to provide a trustworthy transaction record and support a customer in providing evidence of source of funds, recipients of funds, etc., to law enforcement agencies.  After all, banks enjoy much higher trustworthiness for carefully recording and not tampering with customer transactions than the average customer herself.  But trustworthy, tamper-proof recording that holds up to the scrutiny of law enforcement agencies is no longer out of reach for common customers: Cryptographically secured immutable document structures \citep{haber1990time} have been popularized and made publicly available by blockchain and peer-to-peer systems \citep{benet2014ipfs}. Distilling money into digital cash rather than taking bank money and its stored bank ledgers as the given, the customer who is ultimately responsible for abiding by the law can delegate secure and tamper-proof recording to any trustworthy party or system, without necessarily giving the bank that performs the transfers  access to that information and thus without making the bank co-culpable in case the customer does something fishy.  

Banks are likely to remain the go-to agents as custodians of digital cash and/or tamper-proof record keeping for their customers, since they can provide multiple economic benefits for the customer, even without having to rely on the de-facto legal monopoly to access private and confidential customer information they have been given by KYC, AML and CTF legislation.  In particular, they can still borrow from their customers by offering them deposit accounts, in addition to digital cash storage accounts that are fully backed by central bank reserves.  And they can continue issuing their own bank money, providing their signature expertise of creditworthiness assessment and aggregated credit risk management, and fullfilling their core economic role: Providing inside money (credit money) for facilitating productive economic activities---bank loans.  

Digital cash is about providing direct, equitable, private access to 
secure (default-free, for any amount) fiat money, in analogy to physical cash.  These supplement loans to a bank in the form of deposit accounts, not replace them. 
 
\section{Digital cash in the real world}

Does digital cash exist? Not in its ideal form, where it corresponds to physical cash ``on speed'': secure (no counterparty risk), fully fungible (not discernible from other people's money), directly controlled (no requirement of intermediary), transferred privately (no built-in disclosure to a third party), very fast over any (geographic or jurisdictional) distance (unless explicitly limited according to a specific policy).

Arguably the most important aspect of digital cash is its security, its owner being protected against selective default by a counterparty.  Modern bank money provides speed and low transaction costs, but is typically not secure since a private bank may default without the entire banking system defaulting.  Thus there is a risk that somebody's money can suddenly be worth a lot less than somebody else's, even if they had the same amount beforehand.\footnote{This is in contrast to severe inflation or the collapse of a central bank, where the value of \emph{everybody's} money is devalued equally, rather than selectively.} 

This is not a theoretical risk.  As we have noted previously, fractional reserve banking has essentially two equilibria. Most of the time it works as desired, but occasionally it goes into a bank run phase where customers run to try to extract as much of their deposits before they are depleted and the bank defaults.  Such defaults have been frequent prior to the advent of central-bank fiat money; they are still quite common, most recently in connection with the global financial crisis of 2008, in the aftermath of which a number of banks, including Danish ones, defaulted. 

Some countries offer de facto secure bank money to individuals and companies. For example, the Postal Bank of Japan is majority-owned by the State of Japan, which essentially extends the State's non-default guarantee to the Postal Bank and thus to its depositors, for any amount of money deposited.  PostFinance in Switzerland does not issue credit money; it is required to have full reserves for its deposits in the form of cash reservers at the Swiss National Bank and extremely high quality Swiss government bonds, which guarantees that deposits are fully backed by central bank money and the full faith and trust in the Swiss government and its ability to collect taxes.

These \emph{de facto} full-reserve banks safely \emph{store} money for their customers rather than \emph{borrow} it since they don't create credit money that gets the same status as the money in their customers' deposit accounts.  
There is little historical evidence to suggest that access by individuals and businesses to secure (fully backed) digital money carries the risk of outcompeting private banks and causing a sudden outflow of deposits.  To the contrary, private banks have proven to do quite well in competition with secure state banks.  For example, Denmark does no longer have a bank that provides secure storage of digital money to individuals and non-bank businesses for nontrivial amounts of money.  Only the Danish banks themselves have that access. 

\section{The need for transactional money}


Many discussions of digital forms of money focus on transfers, that is reliably getting money from A to B.  Secure, reliable and efficient transfers are key for the fundamental functionality of money: if you cannot transfer it, it cannot serve as a medium of exchange. 
Despite vast numbers of people still being without any access to digital money or only to semi-automated (and thus costly) ways of effecting transfers, this is arguably a largely solved problem with \emph{bank money}. In many parts of the world, it can be done conveniently by credit card, debit card, app-based (mobile pay), account-to-account bank transfers (via SEPA, SWIFT, Fedwire messaging combined with interbank credits, clearing houses and central bank settlement), or non-bank account transfers (in centralized database systems managed by e-money account providers such as WeChat, Alipay, Revolut); ever more, faster and cheaper transfer solutions are arriving. So, is ``digital money'' a solved problem?  

Providing secure execution of individual transfers in isolation of each other misses the key challenge.  Money transfers occur invariably in connection with other events; they are payments \emph{for something}.  Any theory of money is based on economic resource \emph{exchanges}: money is not a medium of transfer, but a medium of exchange.  

In Section~\ref{exchange-theory} the primitive operation is a two-party exchange of (economic) resources, upon which everything else is built.  
Such an exchange comprises \emph{two} transfers, one from A to B and the other from B to A. Such an exchange must be an \emph{atomic transaction}: One transfer happens \emph{if and only if} the other transfer happens.  Putting this contrapositively, if A doesn't transfer her resource to B, then B does not transfer his resource to A either; or he \emph{assuredly} gets it back if he has already done so.  

Efficiently transferrable digital money is not enough; digital money needs to be efficiently \emph{transactional} since its very purpose is facilitating exchanges, atomic transactions comprising multiple transfers where at least one of them is a money transfer and the other ones will typically be transfers of \emph{other} economic resources, typically physical goods and services.  The fact that there are still vast amounts of losses due to one party delivering something and the other one not paying for it or the other way round suggests that this is \emph{not} a solved problem.


\section{Digital contracts}

A \emph{contract} is an agreement between at least two distinct agents that commits them to a course of action involving \emph{multiple} events such as transfers of money and other resources.\footnote{We refrain from a legal or historical analysis of the concept of contract since we only want to highlight the importance of treating a set of events \emph{collectively}, not only each event in isolation.} A \emph{digital contract} is a formally specified contract where both data and logic are rigorously specified with mathematical or programmatic precision.\footnote{Not to be confused with natural language document templates where only the data and/or the revision and signature process is digitalized.}

An exchange contract is a simple kind of bilateral contract requiring two transfers of resources, where one agent gives something to another in return for receiving something of comparable value from the other.  In Section~\ref{exchange-theory} we have seen a number of more complex contracts arise, in particular financial contracts (money-now for money-later) such as loans, with and without collateralization.  In general, a contract can involve a considerable number of resources, events and agents with obligations, permissions and prohibitions whose conditions stretch over a long period of time. Examples are  bonds, insurance and pension contracts, or complex commercial contracts such as the staged delivery of a cement factory involving production, transportation, trading, inspection of goods and processes, etc, with multiple options, penalties, etc.  In general, a contract may specify many alternative \emph{acceptable executions} (sequences of events) of varying lengths.   

\subsection{Managing contracts}

A contract is a passive object, analogous to a rule book or protocol.  A contract executes itself as much as the paper it is written on: Not at all.  Apart from the contract parties the execution of a contract draws on additional agents.
\begin{itemize}
\item Contract managers: virtual or real third parties the contract parties have agreed on that inform the contract parties of contract events and determine whether actions are in accordance with the contract or not.  These often take the form of a virtual (conceptual) third party consisting of the contract parties themselves communicating with each other, plus a contractually specified court to resolve disagreements.
\item Transaction managers: real or virtual third parties that guarantee that resource exchanges happen atomically, as in payment-versus-delivery in financial instrument trading.  Typically this is done by banks, notaries, central counterparties, security depositories or similar that collectively serve as escrow agents and guarantors of atomic exchange execution.
\item Resource managers: real or virtual third parties that are used to determine legal ownership of certain resources and authoritatively perform transfers of these resources.  Examples are banks for (bank) money, public authorities for land registries, central security depositories for securities.
\end{itemize}

A digital contract is a formal specification of a contract as rule book or protocol, analogous to a (behavioral) type or an interface specification in a programming language; in particular, it does not execute itself.

A \emph{contract manager} for a digital contract is a mechanism\footnote{hardware/software system} that monitors and validates that a contract action such as a transfer is in accordance with what the contract allows or stipulates.  It signals when a contract is successfully completed, manifestly breached (cannot be completed to a successful sequence of events) or is still live (neither terminated nor manifestly breached) and in which state it is.  

A \emph{transaction manager} for a (sub)contract (including but not limited to exchange contracts such as payment-versus-delivery) is a mechanism that \emph{guarantees} that 
\begin{itemize}
\item either an acceptable path in accordance with the mathematical semantics of the digital contract is executed to completion (commit);
\item or, if it does not complete, all resource transfers already performed are effectively undone or compensated for such that the net effect of the failed contract execution is equivalent to no resource transfers having occurred (abort).
\end{itemize}

A transaction manager for an exchange contract thus guarantees that the two transfers either happen and are finalized (committed) or that none of them have effect (aborted).  This ensures a modicum of \emph{fairness}: something-for-something or nothing-for-nothing, preventing something-for-nothing breaches.  A stronger requirement would be that the contract succeeds for the agent(s) who correctly perform(s) their part of the contract even if some counterparty fails to do so.  This is what a central counterparty typically provides in an exchange contract by interposing itself as benevolent man in the middle between the two exchange agents. 

A \emph{resource manager} is a mechanism that keeps track of resource ownership states and effects resource transfers, that is it keeps track of who owns how much money (or another resource) and both performs and validates a transfer; see Section~\ref{ownership-states}.
For example, a bank is a resource manager that keeps balances of what agents own in accounts stored on a database system it controls.  Bitcoin, considered as a \emph{virtual} agent, is a resource manager that keeps track of Bitcoin account (UTxO) balances.  Ethereum does the same for Ether account balances.  It also provides a general-purpose programmable platform for anybody who wants to program and run a resource manager or any other program for that matter.  

\subsection{Smart contracts}

The term \emph{smart contract} was introduced by Szabo in a brief note as ``a computerized transaction protocol that executes the terms of a contract''  \citep{szabo1994smart}.  This is a peculiar formulation.  How could a ``computerized transaction protocol'' \emph{execute} what a contract requires of the \emph{contract parties}?  Should such protocol actually make the decisions required of the contract parties, such as whether and which payment or decision to make at which time, decisions that are usually intentionally left to each contract party without predelegation to a (software-defined) third party?  In other words, is it a program that does things on behalf of the contract parties?  Or is ``execute'' meant in the sense of a program that only checks whether the contract parties' actions are consistent with the contract at hand, analogous to a designated referee in a game who is authorized by the players to respond to their moves by rejecting or approving of them, but does not make their moves? Szabo's discussion suggests that he intended the adjective ``smart'' to be ill-defined and open: using computers in connection with contracts somehow.  

The term became popular and is nowadays closely associated with Ethereum \citep{buterin2014next,wood2014ethereum}. An \emph{Ethereum-style smart contract} is a program that controls a specific Ether account; it is essentially a \emph{software-defined account manager}.  It receives a message (method invocation) from the outside or from another smart contract that can contain both data and a designated amount of Ether that is to be transferred from the sender's account to its own account; execution of the code invoked by the message can send data and Ether amounts from its own account to other smart contracts or to Ether accounts that are not controlled by a smart contract.  Execution is deterministic, 
single-threaded and transactional on a global scale: After the receipt of an outside message Ethereum executes all and only the steps of the initiated computation sequentially and either commits the updated data and Ether balances upon successful termination (commit) or rolls all of them back upon failure (abort).  Ethereum serves as combined contract, transaction and resource manager, where a digital contract is expressed as an Ethereum program. Such an Ethereum-style smart contract thus amalgamates and interleaves contract (the rules), contract management, transaction management and resource management in its code.  This makes for a very simple (single-threaded!), but also low-level and dangerous programming model with direct access to the agents' accounts that are managed by smart contracts, where contracts and their management are conflated.  Additionally, they incur extreme computational costs to implement Ethereum as an open, Sybil-attack resistant, Byzantine fault-tolerant and globally synchronizing replicated state machine.

Ethereum-style smart contracts or digital contracts managed by a separate contract manager do not require a blockchain system.  They could straightforwardly be offered by banks as programmable bank accounts, with more out-of-the-box privacy for participating agents.\footnote{Banking regulation may make this presently difficult and may make banks co-culpable in case customers shoot themselves in the foot or get robbed by bad smart contracts.  This highlights the ambiguous benefits of the decentralized, anonymous governance of permissionless blockchain systems, where there is no protection and no responsibility allocated in case of intentionally or accidentally bad smart contracts.}

\subsection{Separating contract, control and settlement}

Separating digital contracts from their management (control) and resource transfer effects (settlement) has advantages: contracts can be expressed declaratively, close to the compositional structure of paper contracts; they can be analyzed automatically, independently of how they are managed \citep{HLM2020}; and certain operational aspects such as collateral management can be allocated to the contract manager without polluting the specification of the financial contract itself \citep{egelund2017automated}.

Additionally, separating contract management from resource management has multiple advantages compared to monolithic blockchain systems that work by hosting resource ownership, data, and programs manipulating them on the same platform.  For example, resource management can be delegated to existing resource managers such as the banking system (for transfers settled with ``real'' bank money rather than proxy money/IOUs such as stablecoins) and existing central security depositories (for reuse of established, legally sanctioned systems for authoritative security ownership and their transactional services), and contract management can be scaled by massive parallelization since no synchronization between events regarding different contracts is required: a notification by a bond issuer in Russia need not be synchronized with a rental payment by a tenant in Peru.  They not only commute (they have the same effect in either order), they are entirely independent of each other (it does not even make sense to talk about them being ordered).

\section{Contract-backed digital cash}

In Section~\ref{digital-cash-criteria} we started off by stipulating that idealized digital cash be essentially like physical cash, only dematerialized, much more efficient, faster and farther-reaching.  We quickly recognized that digital cash---and arguably any form of electronic money---is deficient without an efficient, reliable, fully digitalized mechanism for atomic economic \emph{exchanges}: money-for-goods, money-for-services, money-now-for-money-later and, more generally transactional (atomically executed) contracts.  A digitalized mechanism for performing one transfer at a time is insufficient.  

We thus stipulate an additional property to the list in Section~\ref{digital-cash-criteria}. \emph{Contract-backed digital cash} is digital cash that is 
\begin{itemize}
\item equipped with an efficient, reliable, secure and confidential mechanism for transactional exchange of resources and, more generally, execution of digital contracts involving not only money transfers, but also other resource transfers and other events.  
\end{itemize}  

Certain aspects of contract-backed digital cash already exist or could exist relatively easily: bank accounts can be equipped with computer programs to provide the same functionality as Ethereum-style smart contracts; distributed ledger technology and blockchain systems provide decentralized governance and smart contract functionality; high-level digital contracts can be used to automatically synthesize, analyze and manage (paper) contracts. 

As long as legislation and regulation is technology neutral and focuses on (monetary and legal) policy and (system) governance without undue bias towards or influence by particular managed server-/cloud-hosted database-backed systems, permissioned distributed ledger technology, permissionless blockchain systems, \emph{etc}, the design space for contract-backed digital cash and other forms of ``programmable money'' remains large. Contrapositively, making implicit unwarranted assumptions such as public decentralized systems necessarily requiring a permissionless blockchain system, token-based money being necessary for transaction privacy, the policy and governance discourse on digital money or retail central-bank digital currency (CBDC) may needlessly be constrained.

\chapter{Public versus private money}
\label{chicago-plan}

\chapterauthor{Fritz Henglein and Christian Olesen} 

	Money has had many different forms and purposes throughout history
	\citep{zarlenga2004lost}. Today public money is often understood as either
	credit given by commercial banks or physical cash, coins and notes, issued by a central bank
	\citep{mcleay2014money}. We discuss a type of money that is neither bank credit, as known today,
	nor notes issued by the central bank. 
	

Today central banks only issue physical cash, which is used less
and less. In the UK, 97\% of all money is held in bank deposits \citep{mcleay2014moneyb}.
If one measures the amount of payments performed with cash, the number
is not as extreme. In 2017, 23\% of all payments done by Danish
households were with cash, which is significantly less than 79\%,
the average for the Euro area \citep{nationalbanken2017}.
Almost all other payments are performed with bank money, which
has the advantage of being convenient since one can perform payments
electronically. 

One of the main problems with (private) bank money is that it is only a promise and thus unsafe: a bank may default on it.
Already in 1987 Tobin argued for public digital money that is safe from bank defaults:
\begin{quote}
``I think the government
should make available to the public a medium with the convenience
of deposits and the safety of currency, essentially currency on deposit,
transferable in any amount by check or other order.'' \citep[p. 172]{tobin1987case}.
\end{quote}
A central bank issued digital
currency (CBDC) could be the solution to this problem. The Bank of
England has published an analysis of the economic impact of a CBDC,
in which they argue, that: ``In short, we imagine a world that implements
Tobin\textquoteright s (1987) proposal for \textquotedblleft deposited
currency accounts.\textquotedblright{} \citep[p. 7]{barrdear2016macroeconomics}. 

A CBDC would therefore partly be introduced to solve the problem of
a safe and convenient form of money available to the public. 

\section{Forms of money}


Government-based money in current circulation can be broadly divided into three subgroups: physical cash, digital currency and
bank money.\footnote{In the following analysis, digital currency will be understood as central-bank digital currency (CBDC).  Note that digital \emph{cash} according to Chapter~\ref{digital-cash} is a particular form of digital currency that closely emulates properties of physical cash.}
Both physical cash and digital currency are sovereign currencies in the sense of being issued by a 
sovereign country's government \citep{mcleay2014money}.

\subsection{Physical cash}

\emph{Physical cash} is a bearer instrument. By bearing the cash you prove
that you own it. Therefore no third party is needed to validate that
you own the money. Physical cash is a claim on the central bank.

When paper notes were introduced
by a trusted authority (originally mostly private banks, later mostly central banks), the note was a claim on that authority to redeem
the note upon demand by the bearer for a specified quantity of a commodity such as gold.
In modern fiat currency systems, however, a claim has no operational effect:
It is a claim to get the claim back when redeeming it.  In other words,
it is effectively irredeemable. 

Physical cash note or coin does not bear interest. A note of 100 kr.~is still 100
kr.~ after a period of time. 

\subsection{Digital currency}

\emph{Digital currency} is an irredeemable claim on the central bank and thus an asset, like physical cash. 
It is kept in an electronic account system at the central bank.
In order to prove that one owns it and thus can transfer it to somebody else, it requires proof of 
identity (ownership).  Transfers are made by sending a payment instruction to the central bank that is authenticated to have been
generated by the account's owner.  

Digital currency can be interest bearing.  That is, the central bank can transfer additional money from its account into
the account holders accounts.
Thus 100 kr.~of digital currency can, after a period of time, turn into 102 kr. The ability to
design interest directly into digital currency is a significant quality, one that separates it from physical cash. 

\subsection{Bank money}

\emph{Bank money} is a claim against a certain licensed economic agent, a commercial bank, that can be exchanged at a fixed one-to-one exchange rate for a 
claim against another commercial bank or the central bank.  
It is also also called credit money since banks can create money \emph{ex nihilo} and transfer it to a bank account in their own or another bank.  Bank money constitutes most
of the money circulating in modern economies. In the UK 97\% of the money is held
in bank deposits \citep{mcleay2014money}. 

By placing money in a bank deposit economic agents effectively
lend money to the commercial bank. Commercial banks then resell this
credit, along with the credit money they manufacture themselves, to third parties who pay a risk premium for receiving it. 
That is, banks invest their customers' deposits together with the credit money they manufacture themselves. 
As long as economic agents
in the economy trust the commercial banks' IOUs and do not \emph{actually} transfer or redeem them in large numbers,
the financial system is stable. But suppose there arises mistrust in a
commercial banks IOUs, which would lead people to pull their bank
deposits. The commercial bank wouldn't be able to pay them out(redeem) them instantaneously, though,
which would induce even more people to rush to pull their bank deposits, resulting in a bank run.

\subsection{Stablecoins}

Stablecoins are cryptocurrencies hosted on blockchain systems that are designed to retain a fixed exchange rate with respect to a sovereign fiat currency such as USD or EUR.  Stablecoins are operated outside the conventional IT platforms controlled exclusively by established banks.   

We exclude other cryptocurrencies or -assets for the sole reason that our investigation is about digital forms of fiat money whose purpose is to be an index of economic activity in the sense of being reliably neither deflationary nor (very) inflationary relative to the amount of goods and services produced in an economy that one can purchase with one unit of the currency. 

\section{How money is created}

Physical cash and digital currency are created when the central
bank issues more of it. 
In contrast to this, bank money is simply an IOU (promissory note) to redeem it for central bank money that a bank can create out of thin air. The reason that an IOU by the bank is
considered money, is that economic agents in general trust bank IOUs
\citep{mcleay2014money}, which is not the case for private IOUs
\citep{kiyotaki2001evil}. 

A new bank IOU observably becomes money when the bank transfers it to somebody else and it thus gets into circulation.
This usually happens when a bank transfers it to a customer's account when entering into a loan agreement or when
purchasing a security.
Following the same reasoning, bank money is ``destroyed'' when customers
repay their loans or repurchase the security. Money can therefore be
created and destroyed by commercial banks \citep{mcleay2014moneyb}.

\subsubsection{What determines the amount of money in the economy?}

Commercial banks cannot create new bank deposits without limits. The
primary mechanism, that determines the amount of bank deposits created are interest rates\footnote{The interest rate can be seen as the price on a loan or an IOU.} combined with capital and reserve requirements the bank has to satisfy.
If there is a high interest rate on bank deposits, then demand for
bank deposits increases, which causes bank deposits to increase
\citep{mcleay2014moneyb}. 

\section{Key questions}

There are several central questions regarding the issuance of digital
money. Above we assumed that the Central Bank would issue digital currency.
This need not be the case. There are several companies 
developing digital stablecoins, which can be a digital currency
that holds it value against a sovereign currency. These stablecoins
would not necessarily be issued by the Central Bank. Therefore the
first question (A) concerns who should issue digital currency. Today almost
all money in the economy is electronic bank money, which results in
the commercial banks having control over a significant part of the
money supply in the economy. How can we create money that does not carry
the risks that credit money carries while still making it convenient
to use? That is the second question (B). In this case it would not
necessarily be the banks that should handle the payment structure
which begs the question, as to who should then (C)? Lastly there are
significant economical advantages with digital money based on digital
contracts. These will be discussed in the last part of this paper
(D). 

\subsection{Who should issue digital money?}

On paper many different operators can issue digital money: 
\begin{itemize}
	\item A financial institution (locally) 
	\item A financial institution consortium with near global reach 
	\item A central bank or consortium of central banks 
	\item A FinTech Company or consortium of Fintech Companies
\end{itemize}
The best issuer probably depends on the objective. 

If the objective is in isolation to issue safe ``digital
cash for people and businesses'' the central bank
is arguably the best issuer since it provides greatest protection from bank default risk. 
If the objective is to offer access to digital
money in a range of currencies with interoperability and without central
points of failure without actually issuing any new money, the best candidates are likely agile 
fintech companies that can quickly connect consumers and businesses to new and rapidly evolving
technical platforms, such as distributed ledger and blockchain technologies, 
for fully digitalized transfers, economic exchanges and 
contract execution. 

This would certainly be the case if such a system should encompass
the entire world and thereby include an objective of effectivization
of monetary systems in the less developed world.

It is unlikely that banks and other prime financial institutions are interested in
offering digital currency to the public as this could undermine their \emph{de facto} monopoly on issuing
credit money and on digital transfers via bank money transfers.
Indeed, the security of digital currency, when not only available to banks but also 
to ordinary individuals and business, is used as an argument \emph{against} making it widely available.
In cases of severe economic, financial and political uncertainty clients could move
their money into the safe digital currency system, exhausting banks' inherently limited reserves 
in a fractional reserve system and causing a bank run and domino-like collapse of the financial system.  


\subsection{How to create safe and convenient digital money without letting	the central bank be the issuer}

Money is supposed to create a fair and just society \citep{zarlenga2004lost}.
We want to create money that is both safe and convenient. Today many
use credit from fractional reserve banks\footnote{Today all banks are fractional reserve banks. With this it is meant
	that commercial banks lend out more money than their reserves. That
	is their liabilities are greater than their reserves. This creates
	risk of default and thereby the cancellation of the credit by the given
	bank. } as money, which carries significant risks. Therefore the government
insures credit money issued by banks, which, however, constitutes a 
moral hazard \citep{tobin1987case}. By depositing money in a full
reserve bank, deposits are considerably safer from bank default. 
Indeed, requiring full-reserve banking was proposed by a group of
economists in the 1930's as the \emph{Chicago Plan} in the USA.

\section{Fractional reserve banking}

The currently prevalent monetary system is a three-tier fractional reserve banking system with the following architecture.  
\begin{itemize}
\item At the bottom, it consists of a single \emph{central bank} that has the \emph{exclusive} privilege and obligation to issue fiat money, a particular \emph{currency} such as DKK.  It produces two forms of fiat money: token-based physical cash and account-based digital currency.  Every agent can own physical cash.  Only banks and a few other organizations can own digital currency.  The central bank acts as authoritative registrar of currency ownership: it maintains an account balance for each agent that is allowed to have an account at the central bank.  

\item Ordinary banks constitute the middle tier.  The central bank maintains bank accounts for the ordinary banks \emph{only}. Their balances represent fiat money; this is \emph{digital} currency since the account balances are kept in computer files only.
Ordinary banks are companies, with private owners and the possibility of defaulting on debt, that is not redeeming IOUs it has issued, and going bankrupt.
\item All other agents constitute the top tier.  Ordinary banks maintain bank accounts for such agents.  
The balances of \emph{sight deposit accounts} represent running \emph{IOUs} issued by the bank that can, in principle, be redeemed for 
central bank money.  They are \emph{not} fiat money.  They are IOUs where only a tiny fraction is backed by \emph{reserves} the bank keeps in its account with the central bank. In practice, the bank has to account for having enough resources consisting of central bank reserves, owners' equity and equity-like capital, future loan income and other assets (buildings, gold, etc) to make it plausible that it has enough value that can, if needed, be liquidated sufficiently quickly to fiat money to cover customers' claim redemptions.  Since \emph{both} equity and loan income and many of the other assets are, in turn, IOUs rather than being physical resources or bona-fide fiat money, there is a danger of indirectly including the bank's own IOUs multiple times.  
\end{itemize}

\subsection{Bank accounts as revolving short-maturity zero-coupon bonds}

In Denmark there are no government-guaranteed banks any more.  So all ordinary banks are allowed to, can go and some have gone bankrupt individually and independently of each other.  This makes everybody having a transaction account in such a bank effectively owner of revolving bank-issued \emph{zero-coupon bonds} (ZCBs) with daily maturity. Every day, the bank redeems the ZCBs issued the day before and offers new ZCBs at the same rate as the day before (unless an interest rate change takes effect that day).  If the customer does not do anything she automatically buys these new one-day ZCBs for all the money from the bonds redeemed that day. The day the bank may go bankrupt such bonds are worth drastically less than the day before.  

If the risk and effect of default is not mitigated by a trustworthy government guarantee, a transaction account is arguably a rather sophisticated financial product.  It is striking that ordinary people only have the choice between physical cash and revolving bank-issued zero-coupon bonds with daily maturity as the simplest and most wide-spread forms of keeping money.  
Completely eliminating physical cash would leave only revolving bank-issued zero-coupon bonds with daily maturity.

Since 2010 banks in the EU have had to contribute to a collective \emph{deposit insurance scheme} that insures customers' revolving bank-issued zero-coupon bonds up to EUR 100,000 per customer against default of the bank, the bond issuer.  That means the \emph{simplest} form of money, apart from cash, consists of revolving bank-issued zero-coupon bonds with insurance against issuer default, but limited to EUR 100,000 per bond holder across all bonds issued by the same bank.   
Amounts above EUR 100,000 are not insured.  Ordinary customers and even small companies often hold more money in their bank accounts, and spreading that money across sufficiently many different banks becomes an important discipline in itself, even though that, by itself, has no productive value to the bank account owners or the economy at large.

Bonds that must be refinanced often are generally considered a potential threat to the stability of the financial system: What if not enough of the redeemed amount is reinvested to buy the new bonds?  

The popular F1 mortgage-backed bonds widely used to fund real-estate purchases in Denmark need to be refinanced \emph{annually}.  The concern over such frequent refinancing auctions potentially failing has led to regulation that the bonds be convertible to permanent bonds with a high floating interest rate instead of having to be redeemed.  In this fashion, the bond issuers are guaranteed against the risk of sudden, massive withdrawal of money needed for full refinancing of the revolving bonds, and the bond holders are protected from the bond issuers escaping making any payment at all by defaulting.  
  
Curiously, financial policy regarding \emph{daily} refinancing of commercial ZCBs seems to favor letting the bond issuer default without compensation above EUR 100,000 to the bond holders.  To minimize the risk to financial stability (but not to individual bond holders), bond holders are effectively \emph{forced} to reinvest their daily bond redemptions in new bonds since they cannot do much else with their money.  They cannot realistically go into cash, only buy other IOUs, also with a counterparty default risk.

\subsection{Should ordinary people be allowed to own digital currency?}

The risk to financial stability is sometimes given as a key argument against making digital currency available to nonbank agents.  The following counterarguments or alternative financial stability management mechanisms can be considered, however. 

\begin{itemize}

\item State-owned banks with \emph{de facto} zero default risk on their deposit accounts have existed and do exist to this day in a number of countries, side by side with private banks.  In state-owned banks, on-demand deposit accounts are effectively fully insured against default and thus behave like account-based digital currency accessible by nonbank agents.  There is little historical evidence to suggest that state-owned banks or full-reserve banks must be banned to make private commercial banks competitive or to ensure systemic financial stability.  

\item Bank accounts correspond to revolving zero-coupon bonds with extremely short-term maturity.  They could be equipped with aggressive reserve requirements corresponding to their maturity horizons.  On-demand accounts (transaction accounts) mature conceptually daily and could require essentially full or almost full reserves.  Full reserve banking has been proposed in the US in the 1940s to prevent financial instability due to the danger of devastating, cascading debt redemptions inherent in fractional reserve banking.  It has been analyzed quantitatively in the context of a contemporary central bank fiat money system and found to be a macroeconomically attractive alternative to the present fractional reserve system; see below.

\item Analogous to convertibility of short-term mortgage bonds to permanent bonds, on-demand accounts could be equipped with a similar conversion policy so that shocks from sudden, massive account withdrawals are prevented.  Banks with only fractional reserves  can then avoid sudden withdrawal of massive amounts of bank money and be kept from---and indeed compelled not to enter into---bankruptcy, which in turn gives account owners the assurance that they can receive their balances as a stream of payments rather than being completely reneged any repayment above EUR 100,000.

\end{itemize}

A fundamental question emerging from this is: Why should ordinary people be denied the benefits of money as an asset without counter-party default risk---something that is manifest, analogous to a bar of gold or a hard-to-forge bank note---just in an even more secure digital form?  Why should they, who are arguably least trained in sophisticated counterparty credit risk assessment, not have access to digital fiat money, but be relegated to lending it  to banks in exchange for bank IOUs with default risk, while banks, the specialists in the discipline of credit risk assessment, have access to safe digital fiat money?  This seems like a reversal of roles.

\section{Fractional reserve banking versus full reserve banking}

The Chicago plan is an attempt to give the government monopoly to
issue money. It is a proposal for full reserve banking,
in which banks must not create money \textit{ex nihilo}, but 
must borrow it from the central bank and their own depositors.  They
cannot simply create new deposits \citep{benes2012chicago}.
Accordingly, this would be a financial
system with 100 percent equity financed enterprise \citep{simons1946debt}
or a model where banks can finance their enterprise through a combination
of equity and non-monetary liabilities \citep{benes2012chicago}.
Simons' and Fischer' main concern with the fractional reserve system
was the instability of credit \citep{benes2012chicago}. As explained
above, almost all money in modern economies is essentially credit issued by banks \citep{mcleay2014moneyb}.

\citet{Fisher1936} presents four claims for the Chicago Plan:
\begin{enumerate}
	\item Significantly reduce business cycle volatility
	\item Eliminate bank runs
	\item Large reduction in the levels of public debt
	\item Large reduction in the levels of private debt
\end{enumerate}
\citet{benes2012chicago} find support for all the four claims for
the Chicago Plan using a DSGE model of the US economy. The four
claims are explained in turn below. 

\subsection{Significantly reduce business cycle volatility}

Fischer and many of his contemporaries believed that the amount
of bank credit issued was the primary source of business cycles \citep{benes2012chicago}.
This poses several problems. Since bank credit doesn't necessarily
depend on variables of the real economy but on the banks' willingness
to supply credit, and changes in the supply of bank credit influences
variables of the real economy, banks have great power to influence
the real economy. In our financial system where almost
all money is bank money the commercial banks essentially control the total supply of
money. The quantity of money and the quantity of credit is therefore
dependent on each other \citep{benes2012chicago}. Under the Chicago
Plan this would not be the case, since the central bank would be able
to control the amount of money in circulation and banks the amount of credit (loans given). This
would cause banks to become intermediaries that can only provide credit
when they have obtained funding for it. Therefore a change in the
banks' willingness to provide credit would not be able to cause business
cycle volatility the same way as it can in the present financial system
\citep{benes2012chicago}. 

\subsection{Eliminate bank runs}

In a financial system of fully reserve-backed bank deposits the risk
of bank runs would be virtually zero. Then banks would not have to
worry about sudden decreases in their bank deposits. For the possibility
of bank runs to be eliminated two conditions must apply. The banks'
monetary liabilities must be backed by government issued money. Since
banks are not allowed to create their own money, this will be true.
Secondly the banks must fund their credit by non-monetary liabilities
whose value is not threatened by sudden changes in demand. This is
because the liabilities funding the banks' activities otherwise could
function as near-monies \citep{benes2012chicago}. In order to achieve
that this does not happen one can do different things. Banks can for example
fund their credit by a combination of government equity and and loans \citet{benes2012chicago}. Especially important 
and difficult to enforce 
is that banks not fund their credit with private debt instruments
\citep{benes2012chicago}, since these are IOUs without full reserve backing and thus
reintroduce fractional reserve banking.

\subsection{Large reduction in the levels of public debt}

By implementing the Chicago Plan the commercial banks would be required
to borrow reserves from the treasury or central bank in order to fund their liabilities.
This will create a large asset for the government, with a large negative
net debt for the government. Since the government issues fiat money, which cannot be 
redeemed for something else, reserves borrowed by the banks are not government debt, but government
equity \citep{benes2012chicago}. 

\subsection{Large reduction in the levels of private debt}

Since the government has a large net debt after issuing reserves to
the banks, it could buy back much of the private debt from banks with
cancellation of treasury credit in return for banks. This would potentially
result in most private debt being bought by the government. This has
the advantage of creating low debt sustainable balance sheets for
both the private sector and the government \citep{benes2012chicago}. 

\subsection{Additional benefits of the Chicago Plan}

\citet{benes2012chicago} argue for an additional benefit
of the Chicago Plan:  Due to fewer distortions in the financial system
caused by banks creating credit risks, there could be a large steady-state
gain. 

\section{What to learn from the The Chicago Plan}

The benefits of full reserve banking where only default-free public institutions such as the central bank can issue sovereign fiat money are hotly contested.   Full-reserve banking effectively guarantees \emph{all} bank deposits, but requires central bank and government to become not only lender of last resort (for banks), but effectively funder of all credit, and thus to carry the costs of bank defaults instead of depositors and other bank creditors.\footnote{It has been argued they are doing that anyway in fractional reserve banking, even with less control over the credit issued when banks that are ``too big to fail'' or depositors cannot be burdened with the losses in a bank default anyway.}  

With full reserve banking, security and fungibility of digital cash could be provided by bank deposit accounts.  It should be   
noted that security from default does not require that \emph{all} bank accounts be secure, but only \emph{some}, and that non-banks get access to them.  
Banks could for example offer special deposit accounts to individuals and non-bank businesses that are fully backed by central bank reserves and high-quality government bonds without limit on the deposited amount.  Common access to such accounts or to (secure) digital currency is essential to preserving the property of a money as a stable and reliable store of value.   A pension fund should not have to rush to convert 100 million Euros that had been transferred to its bank account one Friday evening into securities, driven by worries that the bank could declare bankruptcy during the weekend.

\chapter{Blockchain and distributed ledger technology}
\label{blockchain}

\chapterauthor{Fritz Henglein}

The term \emph{blockchain} has its origin in the data structure used to store a log of transactions whose state is replicated in an open network of nodes employing a peer-to-peer gossip protocol.  As such, it is a specific data structure for achieving a particular purpose, the atomic transfer of Bitcoin \citep{nakamoto2008bitcoin} between anonymous parties without an appointed trusted organization controlling the process of verifying whether the transaction is valid or not.

In this section we step back and propose general functional characteristics of \emph{blockchain and distributed ledger systems}, including systems not built yet \citep{henglein2018}.  It is analogous to describing the notion of automobile from a functional perspective rather than a particular engine technology.\footnote{If Bitcoin is the Ford Model T of blockchain/DL systems, then the description should include it as an instance, of course, but also diesel-driven vans and self-driving electric-motor driven cars should be captured; they share functional and architectural aspects, but are technologically very different. For example, proof of work is a specific Bitcoin aspect, but does not occur in other blockchain/DL systems, just as spark plugs do not occur in electric cars.}  

\section{A blockchain/distributed ledger ontology}

A blockchain system is a distributed system for managing digital (representations of) resources with certain characteristics and goals.  In this paper we attempt to deconstruct blockchain systems into conceptual components so as to derive a canonical ontology and architectural components for discussing, designing and analyzing blockchain systems.  We begin with a basic ontology.  It is meant to be sufficiently stringent and connotative to facilitate a basic discourse without being too restrictive or formalized to discourage meaningful discussion.

\subsection{Distributed systems}

A \emph{distributed system} is a \emph{network} of \emph{(computer) nodes}, each running some \emph{program} that can receive and send \emph{messages} to/from other nodes and from \emph{client (computer)s} via \emph{network connections}.  Collectively, a distributed system offers a designated \emph{service} to its clients via an \emph{interface}, the kinds of messages it receives and sends.  


A node can be thought of as a stateful \emph{object}: it has an internal \emph{state} that may change and cause messages to be sent as a consequence of receiving messages and of internal activities.  The state is not shared with any other node; sharing needs to be explicitly modeled as a separate node or as a distributed (sub)system in its own right.  
A node is \emph{reactive (event-driven)} if its state only changes as a consequence of messages received; that is, it has no active threads of computation that change its state without being initiated by receiving a message.  A node is \emph{controlled} by an agent (see below), its \emph{node operator}. A distributed system is \emph{(organizationally) centralized}, if all its nodes are controlled by a single agent.  It is \emph{decentralized} if its nodes are controlled by a dynamic group of agents that are not, themselves, controlled by a single agent or colluding to act like a single agent.  A node may suffer a \emph{crash fault} (become non-responsive) or even a \emph{Byzantine fault} (does not follow the agreed-upon protocol).  A distributed system may or may not tolerate crash faults and Byzantine fault, that is still provide its service to a high degree in the presence of crash or even Byzantine faults.  A decentralized system may furthermore be exposed to \emph{Sybil attacks}, where a group of agents takes over or adds a large number of network nodes to compromise the system's service by large-scale Byzantine faults. 

Messages are usually classified into \emph{queries}, \emph{responses} and \emph{commands}.  A query results in a response that is sent by the query receiver to the query sender.  Queries furthermore do not change the state of a node: the same query received multiple times yields the same response if no command is received in between.  Commands generally update the state.  The separation into query/response pairs and commands constitute the basic building blocks of CRUD-based and RESTful \citep{fielding2000architectural} programming. 

\section{Resources, agents, contracts and events}

Blockchain systems deal with (digital representations of) economic events and attendant information exchange.  We find it useful to employ the Resources-Events-Agents (REA) accounting model \citep{mccarthy83}, its basic terminology and its subsequent refinements, extensions with information, location, independent perspective \citep{jacquet2003}, structured contracts \citep{AEHSS2006} and other additions in this paper. 

A \emph{resource} is something physical or ideal that is economically scarce such as money/currencies, assets, physical resources (such as trucks, houses, dog food), property ownership, usage licenses, etc.  A resource may be \emph{unique} (the Guernica painting) or \emph{fungible} (50 dollars, 14 liters of milk, a dozen bagels). The characteristic property of a real-world resource is that it is \emph{hard} to copy cheaply by nature (a truck) or by design (money).\footnote{We may use the term \emph{linear} resource for emphasis to minimize the confusion with uses of ``resource'' in IT such as in RESTful programming, where it corresponds to stored information that can be cheaply copied.}  

A piece of \emph{information} is something physical or digital that is economically plentiful such as a message on a bulletin board, an invoice, a picture, etc. The characteristic property of information is that it is \emph{easy} to copy cheaply.

An \emph{agent} is somebody or something that can \emph{act}: a natural person or organization such as a company or division of such or an informal or formal association, an ERP system generating events, a software robot, an IoT device, etc.  

An \emph{event} is a significant atomic change of the \emph{state of the world} occurring at some point in time; it includes occurrences of the following \emph{actions} performed by agents:
\begin{itemize}
	\item
	a \emph{transfer} of some resource from one agent to another, \eg Alice giving Bob 50 BTC;
	\item
	a \emph{transformation} of some resource into another resource by an agent, \eg Charlie producing a bicycle from all its parts;
	\item
	a \emph{communication} of some information from one agent to another, \eg Bob sending Alice an invoice;
	\item 
	a \emph{conclusion} by some agent of output information from some input information, e.g. Charlie concluding that he has to pay 100 BTC from two open invoices of 50 BTC each.
	\item
	an \emph{observation} by some agent, often called an \emph{oracle} in this context, of some information; \eg Bloomberg stating that the price of IBM stock hit USD 146.1 June 8th, 2018.
\end{itemize}

We can describe the state of the world in terms of statements of \emph{ownership} and \emph{knowledge} and characterize the \emph{effect} actions have on the state of the world. If Alice owns 80 BTC and Bob 20 BTC and Alice transfers 50 BTC to Bob, Alice owns 30 BTC and Bob 70 BTC afterwards.  Notice that the sum of what they own has not changed: the resources in the world before and after the event are the same.  In particular, after sending the 50 BTC she does not own them any more.  This is in contrast to communication: If Alice knows that IBM stock hit USD 146.1 June 8th, 2018, and communicates this information to Bob, \emph{both} Alice and Bob know it afterwards.  The information has been duplicated. 

Not all actions are \emph{valid} in every state of the world.   For example, if Alice's \emph{credit limit} is 0---she cannot borrow anything---she cannot transfer 50 BTC to Charlie after transferring them to Bob above; in particular, she cannot \emph{double-spend} the 50 BTC.  
Similarly, if she doesn't know anything about IBM's stock price, she cannot inform Bob of IBM's stock price as if it were a fact.  Though, if Bloomberg first observes IBM's stock price and then informs Alice of it, she can subsequently inform Bob of it and provide a provenance trail of where the got the information from.  

A \emph{contract} is a specification of which actions a group of agents is permitted, obligated or prohibited from performing at which time, in which order and under which circumstances.  At its core, a contract is a classifier of \emph{collections} of events: Given a (possibly hypothetical) collection of events, it classifies it as either constituting a correct and complete execution or not.  

Using the term contract is motivated by the conventional notion of (paper) contract, but its interpretation as an event collection classifier goes beyond this.  Depending on the setting and compositional structure we may call a specification of such a classifier also a \emph{business rule}, \emph{policy}, \emph{mechanism} or \emph{protocol}.  

For anonymization, authentication, authorization and other reasons we may use \emph{identifiers} where \emph{entities} such as resources, agents, events, contracts, nodes, are required.  Their mapping to actual entities is dynamically managed by an \emph{identity management system}, which may be distributed itself.

\section{Blockchain/distributed ledger system characteristics}

A \emph{blockchain/distributed ledger system } can be characterized by the following properties: 

\begin{description}
	\item[Organizational and technical decentralization:] It is a distributed system whose nodes are controlled by a dynamic group of independent principals (organizations), each of which ideally has the same access to and control of the blockchain system.  
	\item[Tamper-proof shared storage:] It maintains a ground truth of shared facts (consistency), which are furthermore immutable (tamper-proof).
	\item[Forge-proof digital resource management:] It guarantees that (digital representations of) assets (resources) can be stored and transferred, but neither duplicated nor lost. 
\end{description}

Specifically, organizational and technical decentralization involves an open (``permissionless'') or closed (``permissioned'') group of agents (parties, companies) that have equal data access, update and administrative control rights to a peer-to-peer distributed system; in particular, there is no privileged information aggregator and process owner (in particular no cloud hosting provider with privileged insight into the information streams of its users).
Tamper-proof recording of events is a technical guarantee against tampering with recorded information, including deleting it.  Recorded events establish a joint ground truth across all agents.  It therefore includes comprehensive reconciliation of inter-organizational information and establishes dependable provenance of information and resources across arbitrarily long supply chains.  Forge-proof digital resource management means that resources such as fiat money, cryptocurrencies, property rights, licenses, (proxies/digital tokens for) arbitrary physical resources (trucks, components, plants, cement bags, cancer medicine,...) can be reliably stored and transferred digitally.  The system has built-in guarantees against duplication (forging) of resources.   This provides the equivalent of having a purely digital ``original'' certificate of anything of value and establishing who, uniquely, owns it at any given point in time.

Figuratively, a blockchain/DL system is the equivalent of a direct democracy involving many individuals spread over a large geographic area with no leaders or hierarchy, yet such that it collectively behaves reliably like a single organization that stores resources (who owns what), gives largely consistent (the same) answers no matter who amongst its (honest) members is asked, and works effectively even when there are failing, cheating or even actively attacking individuals forming gangs and inventing new identities of individuals (ballot stuffing).

There are various inherent distributed systems trade-offs between consistency, responsiveness, tolerance of network (communication) failures, degrees of resilience to failing, cheating members and to ballot stuffing inside the organization, between privacy and performance.  These preclude a single design of a blockchain/DL system being best at everything.  There are many possible different designs; any one of them is technically complicated, and they all are different under the hood from centralized/cloud-hosted database systems, which most software engineers and developers are trained to deal with.

\section{A brief review of blockchain and distributed ledger systems}

We briefly describe the essential parts of some popular blockchain systems.  

\subsection{Bitcoin}

The Ford Model T of blockchain systems is Bitcoin \citep{nakamoto2008bitcoin}.  It is an open network of self-authenticating replicated state machines collectively maintaining a list of transactions.  Each transaction consists of a number of inputs and outputs, the latter with associated nonnegative amounts of its cryptocurrency, Bitcoin; a transaction is \emph{valid} if each of its inputs refers to an output of a previous transaction that has not been spent (used as input by another transaction), an efficiently checkable predicate holds, and the sum of Bitcoin at its inputs is equal to the sum of Bitcoin amounts the transaction associates with its outputs.  The predicate in question is typically proof of knowledge of the private key corresponding to a public key serving as an anonymous identity (``address'') that a Bitcoin transfer is made to; in other words, it is a digital signature by somebody who knows the private key in question.   Bitcoin employs a decentralized lottery where a node needs to compute a nonce (a winning ticket) to prove itself as legitimate validator of a new block of approximately 1,000 transactions that it then sends to other nodes.  Nodes are incentivized economically to extend the longest sequence of already validated blocks by receiving Bitcoin for block validation so that the nodes converge on a single chain of blocks.

\subsection{Ethereum}

Ethereum \citep{buterin2013whitepaper} employs Bitcoin-like blockchain storage.  Whereas in Bitcoin the behavior of an agent controlling an address is opaque, in Ethereum it is possible to tie an address to a reactive object with explicit, immutable code (``smart contract'') that receives and sends messages containing data (information) and its own cryptocurrency, Ether. 

Both systems are intended to work in an open, anarchistic setting: any node and any address can participate; network nodes and users controlling an address authenticate themselves.  No external authentication or permission from any authority is required.  
Byzantine and Sybil attacks are countermanded by making it computationally extremely expensive to construct \emph{any} valid chain of blocks and rewarding convergence on a single chain of blocks, which is then considered the ``real'' one.  

\subsection{Corda, Fabric and other distributed ledger systems}

Blockchain subsequently received much attention as a decentralized platform for establishing (normative) consensus on a sequence of events amongst \emph{authenticated} nodes and agents, especially in the financial sector where resources (money, bonds, etc) are essentially purely digital.  Since payment for operating a node and recourse in case of cheating can be handled outside the blockchain system, such distributed ledger systems typically employ classical distributed systems techniques, without cryptocurrencies for incentivizing correct behavior and without requiring payment  for each computation step. For example, Corda \citep{brown2016corda2} employs an architecture where authenticated nodes send information to each other privately (point-to-point) and only transfers of resources need to be validated.  For this they employ a small set of trusted validator nodes that collectively agree on a total order of all requested resource transfers and then check whether each is valid in the consensus order.  The private messages are, a priori, only available to the sender and receiver; Corda has functionality for cooperatively collecting private messages and validated resource transfers to prove to each other or a third party whether a sequence of events abides by a designated protocol (``flow'') such as a financial contract.  If a node fails or refuses to participate in this phase, this may fail.  

Similar to Corda, Hyperledger Fabric employs an ordering service to globally and totally order functional (deterministic) update requests to the shared state of the world, which are then propagated and applied in that order by all peers in the system to update the state.  In contrast to Corda, but similar to Bitcoin and Ethereum, Fabric nodes are, a priori, replicated state machines that can see all messages amongst any nodes.  (Partial privacy is regained by organizing the network hierarchically: a channel is a Fabric network in its own right, and each such channel participates as a single node in a higher-level network such that all intra-channel messages are kept secret from other channels.)

\chapter{Tokenizing invoice debt} 
\label{tokenization}
	
\chapterauthor{Gert Sylvest}
	
Access to finance is one of the key challenges for Small and Medium Enterprises (SMEs) globally. With \$9 trillion held up in outstanding receivables \citep{auboin2016}, and with payment terms mostly being dictated by larger companies, SMEs face significant liquidity gaps. According to world bank studies \citep{worldbank_on_smbs}, access to finance is one of the key challenges to SMEs globally, and more than 50\% of all SME requests for trade financing are rejected. According to various estimates, a trade financing gap of  \$1.2-2.6 trillion exists \citep{worldbank_on_smbs}. For SMEs who do have access to finance, interest rates are in general more than 50\% higher compared to larger enterprises. 

The reason for this is largely that SMEs are opaque from a risk estimation perspective: Accessing information about SMEs' solidity, maturity, trade relations, transactions and history is generally a manual undertaking that relies on paper and other unstructured data sources. 



\section{The impact of emergent technology} 

With blockchain a technology has emerged that offers a new approach to addressing the coordination of transactions between multiple parties. It does this:

\begin{itemize}

	\item By offering a shared ledger between the parties that is both persistent and resistant to allowing the interests of any particular party to influence its availability or content.

	\item By offering a means to evaluate contracts on information in the ledger that is transactional in nature, supports multiple parties, is deterministic and transparently records its evaluation to the shared ledger for every participant to see.

\end{itemize}

\section{Tradeshift and MakerDAO} 

In 2018 two companies with Danish roots decided to join forces to address this challenge with blockchain. Tradeshift is a global B2B cloud-based supply chain network and application platform that in 2018 is moving transactions worth approx.~\$500 billion between companies in its network. MakerDAO is the creator of DAI, the first decentralized stablecoin based on the Ethereum blockchain.

The fundamental question to be explored is if blockchain can be used to create a more transparent, liquid and competitive market for receivables financing, and thereby create a new class of accessible and affordable financial options for SMEs globally. 

In this model, receivables and other valuable transactions exchanged between parties on the Tradeshift platform will be recorded on the blockchain in the form of tokenized ``IOU's''. These tokens inherit the properties of the relationship they are born in, such as for example the size of the buyer and supplier, the payment intent of the buyer, and the characteristics of the process it has been involved in (such as 2- and 3-way matching with purchase orders and goods receipt advice). Financiers will compete on access to available tokens, and combine information from the Tradeshift platform with information that financiers have access to, such as insights into specific markets or verticals. 

\section{A prototype system} 

MakerDAO and Tradeshift have collaborated and developed a prototype within this space, whose model is described in broad strokes below.

\begin{minipage}{\linewidth}
	\centering
	\vspace{1cm}
	\includegraphics[width=0.8\linewidth]{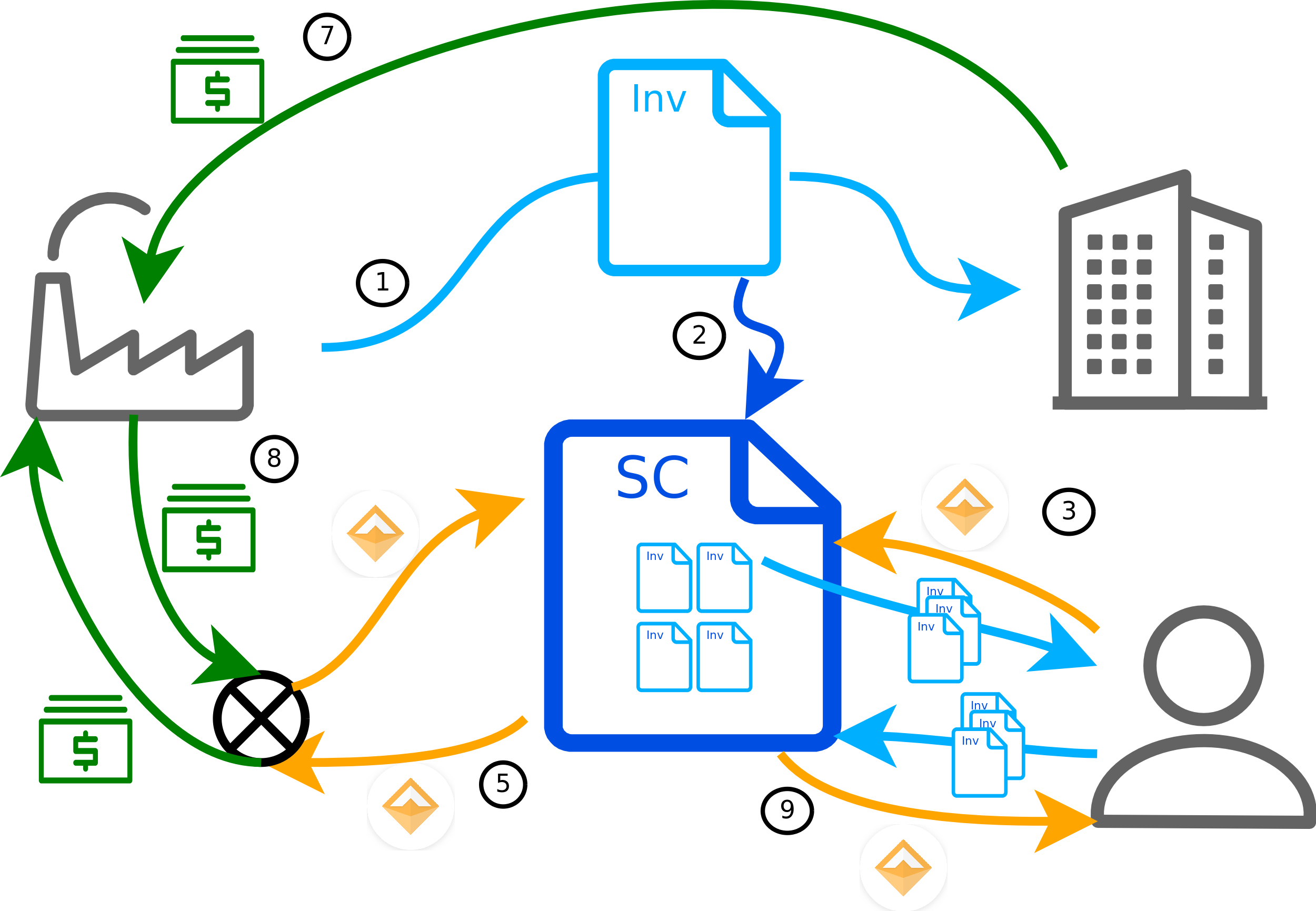}
\end{minipage}

\begin{enumerate}
	\setlength\itemsep{0.05em}
	\item Seller submits an invoice against Buyer and requests financing for it.
	\item A financing smart contract tokenizes the invoice into small parts.
	\item Financiers buy tokens of the IOU for e.g. 0.98 DAI on the dollar.
	\item Once 70\% of the invoice total is covered, the financing completes.
	\item Early payment is made to the Seller through a DAI to dollar off-ramp.
	\item 60 days pass.
	\item The Buyer pays the Seller.
	\item The Seller settles at 1.00 DAI to the dollar through an on-ramp.
	\item The Financiers liquidate their tokens and turn a DAI profit.
\end{enumerate}

While not annotated onto the graphic, it is understood that the smart contracts that represent invoices are annotated in a way that makes financiers able to make financing decisions. E.g. ``this invoice is sent to a Fortune 500 company`` or ``this invoice has received the buyer's explicit acknowledgement.''

\section{Inclusive finance} 

The opportunity is now to include multiple parties including financiers, buyers and sellers in settlement smart contracts, that transparently release early payments and buyer payments to the relevant parties. 

The fundamental challenge for this use case is that no clear regulation exists that allows for companies to directly settle on the blockchain, which in practice means that sellers and buyers have no well-defined way of handling settlement directly on a blockchain, thus requiring a workable bridge to exist between the blockchain and real-world currencies and payment infrastructures.

MakerDAOs distributed DAI Credit System provides a transparent and stable token that allows anyone to represent real-world currency settlements on the blockchain. It does so by establishing a soft peg to multiple currencies where the stable tokens are backed by collateral representing a diverse portfolio of assets. 


The ability to tokenize receivables allows a wide range of financial instruments to be modeled, ranging from peer-to-peer based financing involving small, private funders, to institutional finance and large private hedge funds. 

The ability to create an accessible, liquid and transparent marketplace for receivables holds the potential to drive financial inclusion at an unprecedented scale. But there are even bigger perspectives. 

\section{The vision}

As a move from credit rating based lending to transaction- and relationship-based financing, receivables financing will \emph{de facto} drive a market valuation of reputational attributes of companies that receive financing.

The valuation of such attributes could help eradicate another SME barrier, namely that of access to market: It is clear that sourcing from companies that can demonstrate reputation through transactional and funding histories, confirmed by multiple parties on a shared ledger, will stand a better chance of entering new markets or win tenders over SMEs that are more opaque. 

Secondly, a large volume of receivables diversified over industries, regions and company segments will have properties that are similar to commodities or real-world currencies and could potentially be made to form the backbone of digital currencies that are less volatile than the mostly unbacked cryptocurrencies of today. This could for example be achieved through collateralization of DAI CDPs (collateralized debt positions) with a diversified and replenishable pool of receivables from within this framework. And as regulation evolves, further efficiencies could be obtained if a recurring exchange between digital currencies and fiat currencies could be avoided, and settlements handled directly in regulated, stable and trusted digital currencies. 

\section{The necessity of digital currencies}

The above use case depends on the digital currency in the system holding relevance to the jurisdictions that the parties do business in. The digital currency token must be stable relative to the economies of the seller, the buyer and the financier. This is best achieved by selecting the digital currency such that it is pegged to the fiat currency that the seller and buyer chose to do business in---MakerDAO's DAI leads the pack and is pegged to the US dollar by its internal mechanism.

The system is realizable only through such stable digital currencies. MakerDAO implements this by collateralization of other assets, but it is possible to directly stipulate the value of such digital currency \emph{by fiat}, much in the same way that physical cash holds its value.

\section{Stablecoins: an economic perspective}

As we discussed in the previous sections, stablecoins are assets that maintain a stable price with respect to a target (e.g. USD). There are many approaches to build a stablecoin and in this section we will explore them. From an economic perspective, stablecoins can be subdivided into the following general models:

\begin{itemize}
\item Off-chain-backed coins: these stablecoins are backed with real world assets, such as commodities, precious metals or fiat currencies.
They require trust in at least one centralized entity that stores the real world assets (e.g. a bank that stores dollars). Moreover, these stablecoins require trust in the central party that issued IOUs.
These issues can be mitigated if real world assets are distributed among different institutions and if they are auditable. An example of this type of stablecoins is represented by USD Coin by Circle and Coinbase \citep{usdc}. 
\item On-chain-backed coins: these stablecoins are  backed by other cryptocurrencies. The stability of these coins is influenced by the choice of collateral coins.
Since collaterals are held in a smart contract, transparency and auditability are a strong point of this model. The risk of this stablecoins is the volatility of the underlying collaterals. An example of this type of stablecoins is represented by DAI from MakerDAO \citep{makerdao}.
\item Algorithmic coins: these stablecoins are completely independent from other resources and their supply is algorithmically expanded or reduced in order to maintain price stability.
In particular, an increased demand, such that the price is greater than \$1, causes the system to create new coins, so as to increase supply and make the price go back to the target. Vice versa, bonds are used to remove coins from the market in case price is lower than \$1. The risk of algorithmic stablecoins is that if demand stops growing, they will not be able to maintain their price stability. An example of this type of stablecoins is represented by Basis \citep{basis}.
\end{itemize}

\section{Dimensions of the design space}

Each of the previous general stablecoin models can be characterized more by taking into consideration the following design dimensions:

\begin{itemize}
\item Stability mechanisms: as we discussed in the previous section, it represents the mechanism adopted by the stablecoin to maintain price stability.
\item Transparency: it represents the amount of trust that the users should put in central actors. In the case off-chain-backed coins, the central actor is the one who issued IOUs.
\item Auditability: if collaterals are expected by the stablecoin model, it represents the ability of the users to check their availability.
\item Fallback procedures: it represents the capacity of the stablecoins to protect interests of users in case of a failure of the system.
\item Scalability: it represents the capacity of the stablecoin to support its ecosystem, as number of users and transactions grow.
\item Cost: it represents the cost in terms of fees, required by the stablecoin model to the users, in order to maintain price stability plus the cost of the transactions in the context of a specific implementation of a stablecoin model. If we consider the case of a stablecoin implemented on top of Ethereum, the cost of gas to make a transaction is defined by the market, so it is intrinsic to the underlying technology.
While gas price cannot be influenced easily, stablecoins developers can make a difference if they keep as low as possible the computational complexity of the stablecoin transaction. Indeed, the cost is proportional to the gas price and the computational complexity of the transaction.

\end{itemize}

\section{Beyond stablecoins: tokenized assets}

Tokenized assets are a family of assets to which stablecoins belong. While a stablecoin has the purpose of maintaining price stability and it may represent a real world asset (e.g., dollars), a tokenized asset may represent any sort of asset and there is not necessarily a mechanism to control its price. Together with stablecoins, and more in general with cryptocurrencies, tokenized assets represent a fundamental component of decentralized applications built on smart contracts.
Tokenization is the process of converting an asset into a token that can be stored and traded on a blockchain.
The main asset classes that can be tokenized are:

\begin{itemize}
\item Intangible assets: intangibles are not physical assets. They include patents, copyrights, franchises, goodwill, trademarks, and trade names.
They are usually very hard to evaluate. However, they can be represented by a unique tradable token, gaining a value defined by the market. The ownership of the token represents the ownership of the intangible asset.
\item Fungible assets: a good is fungible when it can be exchanged for another identical good of equal quality and quantity. The most common fungible goods are commodities. Fungible assets are usually backed by a physical resource. Tokenization of these assets allow to trade their ownership instantaneously and with no intermediaries in the process.
\item Non-fungible assets: a good is non-fungible if it is unique and cannot be replaced by another good. A non-fungible token may represents an item of a digital baseball cards collection or an object of an online massive multiplayer game.  
Non-fungible tokens became popular with CryptoKitties \citep{kharifcryptokitties} at the end of 2017.
\end{itemize}

\chapter{Municipal public administration with smart contracts}
\label{municipal-administration}

\chapterauthor{S{\o}ren Debois}

In this section we discuss the potential impact of adoption of programmable digital money, in particular contract-backed digital cash, in the context of Denmark.
Programmable digital money allows access to traditional financial tools, in particular credit money and loans denominated in a stable sovereign fiat currency such as DKK, with the additional benefits of smart digital contracts creating economies and financial applications where digital money transfers are assuredly made according to a certain protocol, and with blockchain and distributed ledger technologies providing a robust,  efficient, decentralized platform for seamlessly transferring money and enforcing digital contracts without manual processing and privacy leaks to outside parties.

According to the European Payments Council, at the end of 2017 approximately 80\% of money transfers in Denmark were already digital \citep{digitalpayments}. Moreover, Denmark demonstrated to be a digital front-runner in Europe by the early adoption of digital signatures, which achieved their breakthrough with NemID in 2010. Consumers are used to paying digitally and to use digital services. For these reasons, Denmark is a good candidate for integrating smart digital contracts and blockchain/distributed ledger technologies in many public and private services.
The adoption of digital cash can reduce the number of intermediaries with privileged access to and control of digital payment processes, thus potentially reducing both fees to and outside control by private intermediaries. While traditional digital payment systems do not have explicit fees for consumers, this cost indirectly influences prices of goods and services and/or is rendered in the form of surrendering personal data to unknown third parties.

In the following chapter we discuss a specific use case of these technologies in the context of the Danish public sector. 

\section{Municipal government decisions as smart contracts}

In this section, we argue that a smart contract platform with an associated
sovereign digital currency has the potential to revolutionize 
municipal government. By formalizing the core processes of
municipal government for social decisions as smart contracts, we may \emph{reduce the cost} of
making such decisions, while \emph{increasing both their quality and
consistency}. The former is obviously desirable, the latter would increase
perceived fairness of municipal governance in Denmark. 

Every year municipal governments make formal decisions on a very large number
of cases with basis in Serviceloven \citep{serviceloven}
and Aktivloven \citep{aktivloven}. Decisions based on these laws
may have vital economic and social consequences for affected citizens, in both
positive and negative ways, e.g.~granting or denying reimbursement for earnings
lost when caring at home for a sick child may be the difference between keeping
or losing that home. 

%

Both the process of coming to such decisions and the organizational frameworks
around them are very costly, due to the complexity of the laws in
question. Moreover, partly because of this complexity, the process oftentimes
stretches out in time, sometimes to the point where citizens are adversely
affected \citep{john_klausen_hjaelper_2015}. Finally, again because of this
complexity, municipal governments find their decisions reverted by An\-ke\-sty\-rel\-sen
in a large number cases.  A recent study by Ankestyrelsen found
that more than 60\% (!) of municipal government decisions on so-called \S 42
reimbursement claims \emph{that were not appealed} would be overturned or
returned for reconsideration had they
been appealed~\citep{ankestyrelsen_ankestyrelsens_2017}. 

We propose formalizing the decision process as defined by, e.g.,
Serviceloven and the various accompanying guidelines from ministries etc, as a
smart contract running on a platform with an associated, state-backed digital
currency. The key benefits of such an implementation are:
\begin{enumerate}
  \item It guarantees that municipalities, citizens, and Ankestyrelsen all follow
    the correct process, since the smart contract will allow no other behavior. 
  \item This guarantee in turn means that it improves public perception of
    fairness in municipal government decision making: the smart contract
    cannot reasonably be accused of, e.g., unfairly serving cost-saving
    interests of the municipal government. 
  \item In particular, it ensures that municipalities perform timely
    \emph{payouts} once a decision has been made in favor of the citizen. 
  \item It automates the complaint process, allowing higher authorities such as
    An\-ke\-sty\-rel\-sen to \emph{instruct} the smart contract representing the claim
    to either reset to an earlier stage (when returning a decision), or to
    immediately pay out (when making a contrary decision in favor of the
    citizen). 
  \item It radically reduces cost of administrative overhead in municipalities,
    since the procedural steps are now embodied in the contract. 
\end{enumerate}

Some of these benefits are in fact realisable
without a digital currency; in particular, point (1) was investigated
in~\cite{krogsboll2020}. However, 
\emph{without} a built-in digital currency, 
enforcement has no teeth: a municipal government may very well be forced to 
follow steps in a process, but \emph{cannot} in this setting be forced 
by the smart contract
to disburse funds. 

On a longer timescale, conceivably the inverse processes, where municipalities
demand payment or re-payment from citizens could be similarly implemented as
smart contracts, similar to the current situation where Skat, the Danish Tax Authority, may withhold
outstanding taxes or other debts to the government from citizens' benefits. This
would confer similar benefits of avoiding illegally withholding such
benefits~\citep{trine_schultz_ugyldige_2011,trine_schultz_ugyldige_2016}, while
reducing costs. 

\chapter{A Danish national E-Krone with limited usage}
\label{e-krone}

\chapterauthor{Morten C.~Nielsen and Christian Olesen}

A national E-Krone with unlimited usage and available for everybody is most likely not the way forward for the following reasons:
\begin{enumerate}
\item A National currency issued in digital form by the national bank is the same as having a bank account at the Central Bank. 
\item In the case of a bank run caused by severe disturbances to financial stability in an economy, this could create a highly undesirable economic effect, as everybody not covered by a governmental guarantee as far as bank deposits are concerned would move money into the central bank by merely buying the e-currency with cash deposited in commercial banks, creating a so-called bank run. Due to the digital nature of an E-krone this bank run is expected to happen with greater speed than bank runs in an economy with physical cash. 
\footnote{It should be noted in this respect, that retail consumers in Denmark already have a government guarantee and thereby digital currency available through the so-called \emph{indskydergaranti}, which guarantees all deposits by the government and thus are insured by the central bank for any retail deposits up to DKK~750,000. As such, E-Kroner in digital format already exists based on banking infrastructure.}
\item It is not the role of central banks to compete with commercial banks. It is a role of the central bank to act as the bank of the public sector and the gateway between the monetary system and commercial banks. It would be highly undesirable if a central bank, were to undertake the role of a commercial bank in one shape or another, unless this was part of an intentional political agenda, as was suggested by the Chicago plan in the US from the 1930s.
\end{enumerate}

From a technical perspective, it should also be noted, that discussions around the world concerning a platform for national currency are based on using blockchain technology, which, on the one hand, replaces cash in circulation and, on the other hand, has socio-economic benefits that could be desirable from a political perspective.

However, it should also be noted that if Denmark, for example, decided on one set of technologies based on for example HyperLedger and another central bank in another country would settle on the Ethereum blockchain combined with Raiden technology, they would not be compatible and could therefore not be used cross-border. 

The creation and establishment of a national E-currency should perhaps not be completely dismissed for the reasons mentioned above, as the technology available through blockchain technology has some highly interesting macroeconomic possibilities, efficiency increases in the management of a monetary system and potentially significant savings and growth opportunities throughout an economy.

\textbf{We would like to propose the establishment of a Danish E-krone with limited usage or limited size.} 

A national digital and blockchain based E-krone would be established by the central bank of Denmark and offered to the general public outside the existing commercial banking system. Hence banks do not handle payments or make accounts available. 
The E-Krone is held in digital wallets on smart devices and represents a direct obligation of the Central Bank of Denmark.

All private individuals can have an account with the National Bank, with the option of having deposits of up to between DKK~10,000 And DKK~50,000.
The deposits will have a credit rating of AAA+ and will only be used to finance public and semi-public projects. 
An institution must be created to determine how the money can be used and enforce rules and regulations applicable to this new space.
It should be noted that blockchain technology might not be the only or most desirable option as a platform for an E-Krone for two main reasons: The cost of conducting a transaction might be uncertain in the future and because current blockchain solutions have problems handling GDPR regulations. 
Another option might be to look at a Cloud Native Core Banking solution with smart extensions instead. 

\section{Use case}

Say for example that a municipality wants to finance the establishment of the local community assembly hall, based on public and democratic votes, but cannot access funding from either the state or increased municipal taxes, and the project is beyond the legal boundaries of general municipality projects. By having access to deposits in \emph{Danmarks Nationalbank} made by private individuals the municipality could raise funding for the Assembly Hall by covering half of the funding and let loans from individuals from E-Kroner accounts fund the rest. 
The cash flows could be handled by blockchain technology and the loan structure managed by a smart contract, and hence the payment structure works completely independent of the banks. 

Such a system could make it possible to finance public projects, which are not possible today, which are not purely based on economic or social values.
Other examples could be:
\begin{itemize}
\item Re-planting a new local forest
\item Creating a new recreational area in the municipality
\item Nature restoration
\end{itemize}
Also, the system could even be used, as a gateway towards local non-economical collard coins which ties into local projects of high social value with no specific economic return such as elderly visits, playing football the kids and much more.
More dramatic ventures could be co-financing of infrastructure for the bridge project between Germany and Denmark or a traffic harbor tunnel system in Copenhagen.  

Based on the principles of peer-to-peer lending, a project could theoretically be interest-bearing as well, depending on the rules set out for a specific scheme.
A central governing authority should probably be established as a gateway between the individual projects proposed through local municipalities on behalf of citizens and the money they hold in E-Currency.
One of the benefits for the participants is that they will have the possibility of financing projects without paying the banks for handling the payments and the establishment of the funding vehicle.

A governing mechanism for allocating capital to public projects should be based on democratic principles and not be seen and used as an extension to public funding vehicles. Projects that the citizens believe in should go forward and are likely to attract the most significant funding. 
Projects that can't be financed directly through banks can be financed at potentially more competitive interest rates for investors and cheaper financing for the borrowers. 
We suggest a limited scope to start off and then continuously analyzing the use and development of such a new peer-to-peer investment/funding structure.

\section{Additional political and economic benefits}

An additional political advantage could be that economic activities can continue even in the case of severe financial disruptions such as what we saw in 2007/08, where lending stopped because the flow of money from the banking system stopped working.
Further, in a world where global banking systems are interconnected, it could be beneficial from an economic stability point of view to have savings accounts that are not directly part of the banking system, and do not form part of the traditional fractional reserve based lending systems and the risks associated with this,  but represents merely cash holdings in digital form that over time can reduce the need for public guarantees of retail deposits in commercial banks. 

Whether E-Kroner as such should be interest-bearing or not is beyond the scope of the present working paper, but the possibility exists.

A national E-krone with 3 million users in Denmark depositing a maximum of DKK 10,000 would be equivalent to DKK 30 billion and DKK 50,000 equivalent to DKK 150 billion being available to a lending economy that is guaranteed to be without private bank created credit money, and the amount could be increased if desired.
Corporations should not initially participate in this limited edition of Danish E-currency because they are subject to different economic rules and operations compared to regular retail depositors.

The M2 money aggregate in monetary economics is more or less obsolete in advanced economies. A way to handle digital cash could be to divide cash in circulation (not tied to M1 for banks) into two M2 components: M2 physical cash and M2 digital cash (E-Kroner). 
Blockchain technology offers some technical advantages. It provides an indisputable ledger of transactional history that, combined with traceability of the underlying flows of money, can be used to enforce usage of the E- currency in economic sectors that are known to be associated with tax evasion and other activities not compatible with current laws. In other words, the M1 cash (M2 cash) component could over time be reduced in favor of M2 digital cash to limit illegal usage of cash.

An E-Krone can also over time be tied into the payments and receivables of public services like the regular payment of car taxes and students receiving student grants.

In theory, each project could represent economic value with or without interest which could be liquid if allowed to trade on an authorized exchange. Such a situation could, for example, occur if the project is very local but external money from another region is required to finance a project fully.
Take for example the restoration of a Danish beach, with high economic tourist value in the summer months for the local communities. The restoration of the beach could result in significant local revenue increases in its tourist season and could hold economic value, which means that interest could potentially be paid on the loan and attract funds from the general public holding E-kroner.
As there is a need for a direct link with ordinary bank accounts, like what is called "NEM Konto" in Denmark (easy account), the entire system could easily be brought back and scrapped again if necessary.

\section{Technological use case}

Blockchain technology and distribute ledgers represents a very significant change in the way we handle assets and data in the future. Deploying the technology at such a core part of an economy like money could result in the adaptation of such technologies throughout an economy, with significant ad hoc and side benefits for the overall economy going forward. 

\begin{center}
\includegraphics[width=1.0\linewidth]{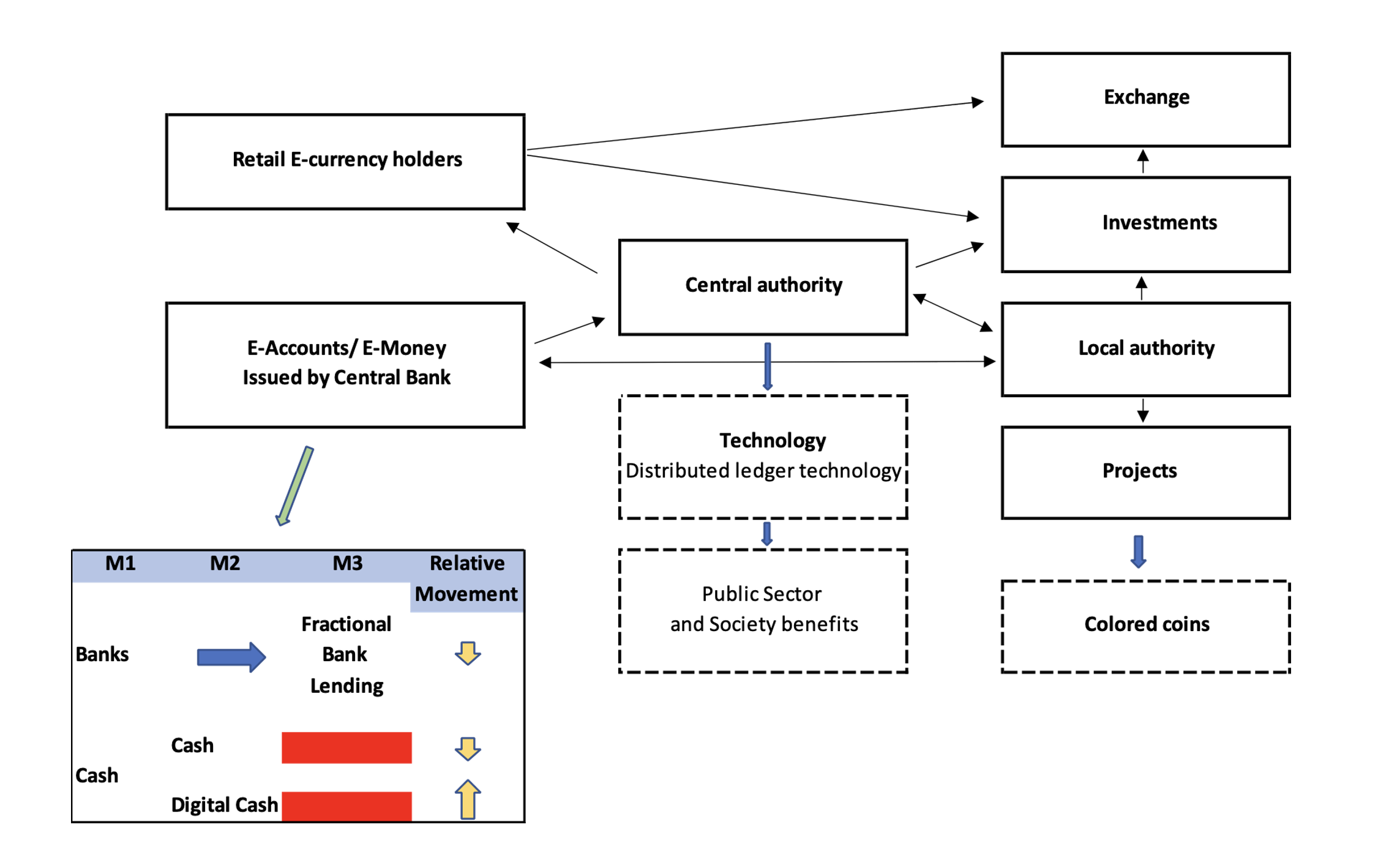}
\end{center}

\bibliographystyle{abbrvnat}
\bibliography{p}

\appendix

\chapter{Proposal: Contract-backed digital cash}

This is the original project proposal.  Only select parts were intended and 
expected to be pursued, which was made possible by external funding of DKK 150,000 awarded by the Danish Innovation Network for Finance IT to each of the academic partners, Copenhagen Business School, IT University of Copenhagen and University of Copenhagen.  

\section{Abstract}

Digital cash is fiat currency stored and transmitted electronically.\footnote{We use the term to encompass central bank digital currency, commercial bank issued deposit money and cryptocurrencies including so-called stablecoins aimed at emulating fiat currencies or achieving low volatility and high liquidity by other means.}  
We stipulate that digital cash is not interesting for performing tracking and analyzing payments in \emph{isolation}---an essentially solved problem---but as an efficient facilitator of provably secure and efficiently executable and analyzable interdependent \emph{complex transactions} such as \emph{formal contracts} involving a multitude of resources, agents, contracts and events, both physical and digital.  

We propose to:
\begin{itemize}
\item identify and analyze relevant \emph{dimensions} of the \emph{design space} for digital cash with an emphasis on its role in complex transactions;
\item analyze the economic potential of blockchain-based payments, contracts and other transactions for Denmark;
\item \emph{inform} the subsequent design of distributed systems for introducing \emph{contract-backed digital cash} in Denmark;
\item and perform 1--3 \emph{proof of concept} experiments to demonstrate and evaluate the designs.
\end{itemize}


\noindent The academic partners are Copenhagen Business School, 
IT University of Copenhagen, 
and University of Copenhagen. 
The industrial partners are Aryze, Deon Digital, MakerDAO, Tradeshift in Phase 1. Other partners are welcome to join in Phase 2.

\section{What is digital cash?}

\emph{Money} is a tradable IOU: A promise by its \emph{issuer} (which may be nature, as in the case of gold) to give its present \emph{owner} something worth the stated amount.  \emph{Fiat money} is money issued by an authorized issuer, such as a central bank, a commercial bank or a computational process (e.g.~computing a small hash code).\footnote{The presentation here is simplified; fine points are relegated to the analysis proper.}  The issuing may involve a reverse IOU, as in a bank loan: a commercial bank creates money out of nothing (an account deposit) and an associated reverse IOU, its repayment requirement (the account owner's debt in the corresponding amount).  Paying off the loan's principal eliminates both IOUs; the deposit money originally created and circulated for a while is destroyed again.  In this fashion, money can be both issued and ``unissued''.  

Money is, for economic purposes, \emph{fungible}: Its \emph{amount} is solely important, not how that amount is split into designated  sub-amounts.\footnote{In practice, such splitting, as in physical cash and certain cryptocoins, is important: It adds complexity (turning one splitting of an amount into another) and admits tracking of the sub-amounts.}  Money is \emph{anonymous}; it is separate from its ownership.\footnote{This does not preclude tracking payments; it only states that the notion of money does not \emph{include} who owns it at any given time.} 

\emph{Cash} is fiat currency in the form of bills and coins with designated amounts issued by a central bank.  They are bearer instruments: Bearing the cash is proof of owning it.  

Proof of ownership of money is necessary and sufficient authorization for a \emph{payment}: transferring its ownership to somebody else. \emph{Transferring} is \emph{linear} (no double spending) and \emph{final} (nonrepudiation): the same bill/coin cannot be transferred twice, and the transfer cannot be regretted (taken back by the previous owner).\footnote{This does not preclude the previous owner having a legal claim to its repayment.}

\emph{Digital cash} is just cash, but with bills and coins as \emph{physical evidence} of owning the IOU replaced by \emph{digital evidence}. As such it should be thought of as analogue to physical cash: it is directly controlled and transferred by its owner without intermediaries; transfers are final, in particular it can be lost and stolen; it can be transferred or loaned to trustees or banks (analogous to putting cash into a bank account); its safe-keeping can be entrusted to others (analogous to keeping valuables in bank vaults or cryptocurrencies).  But its digital form provides a litany of new functionalities, including but not limited to tracking spending (provenance) and what is has be used for (contracts it has been used in), rapidly issuing it (helicopter money), etc.

%
%
%
%
%
%

\section{It is not about transfers (payments)}

Issuing digital cash is actively considered and pursued by a number of central banks.\footnote{People's Bank of China, Bank of England, Bank of Canada, Monetary Authority of Singapore, Sweden's Riksbanken, US Federal Reserve, central banks of Russia and Ukraine; possibly more.}

They seem to focus on payments.  A payment changes account balances, which, abstractly, is a mapping of account identifiers to (money) amounts.  Such mapping may be centralized or distributed in a computer system or physically distributed across the locations (pockets, under mattresses, etc.) where bills and coins are located.

No blockchain\footnote{We use the term blockchain interchangeably with distributed ledger here.} keeping track of payment history is necessary; it is only the balance of such payments that matters.\footnote{The blockchain in bitcoin is an implementation aspect of maintaining account balances: it is tamper-proof, sybil-attack resistant, efficiently checkable \emph{evidence} of the correctness of the account \emph{balances}; it is the account balances that are solely relevant for verifying whether an account owner can transfer a certain amount of bitcoins or not.} For verifying whether any single payment is valid, it is sufficient to keep track of account balances.

Digital cash \emph{accounts} already exist: Commercial banks have digital cash accounts at their national central bank.  
Much of the existing economic analyses of digital cash focus on what happens if not only commercial banks, but all persons, natural and judicial, are allowed to have central bank digital currency: supporting and managing negative interest rates, providing helicopter money, supplanting artifacts and consequences of slow settlement (ownership transfer of assets) such as repo markets and clearing houses, etc.\footnote{See Bank of England's OneBank reports.}

\section{Why conventional payments are efficient}

It is instructive to perform a back-of-the-envelope calculation as to what it takes to process a representative volume of payments in conventional payment system with hierarchical gross settlement.  

As of 2017 there are approximately 100 million Dankort transactions per month, about 1.5 billion transactions per year.  At 1 kB per transaction, they require about 
1.5 TB per year, which fits on a single USB3 disk
costing less than DKK 500.  The transaction volume translates to an average transaction rate of 50 per second, which requires about 500 kb/s throughput, assuming 1 kB data exchange per transaction.\footnote{In comparison, Bitcoin has presently roughly 5 per second, Ether about 3 per second worldwide.} Capping throughput at 10 times the average thus requires only 5 Mb/s bandwidth.  This is the rate at which transactions would need to be streamed to storage for journaling and recovery purposes; a cheap USB3 disk mentioned above provides 100 MB/s throughput or more.  Only the current account balances, not the individual transactions, need to be maintained and queried efficiently to verify payments, that is checking that the account balance is not less than the amount to be transferred.  At approximately 5 million Dankort in circulation this requires maintaining the balance for about the same number of accounts.  This requires less than 20-40 MB storage to support efficient random access and update; this easily fits into RAM of a small PC, indeed it fits inside the level-3 cache.  Assuming reliable point-to-point internet connectivity with 25 ms latency---a not entirely unreasonable assumption---a single early 2000s PC placed almost anywhere on the internet in Denmark could, \emph{ in principle}, support all Dankort transactions in less than 100 ms per transaction.  Even when accounting for security and fail-over (but \emph{not} the complexities of layers upon layers of software and organization and their historical evolution) by multiplying throughput requirement by a factor 100, this suggests that a \emph{very small} and simple distributed system with a moderate amount of redundancy and crash fault recovery is sufficient for providing sub-second electronic payments at the rate of Dankort transactions.


\section{It is about contracts}

A blockchain is sufficient, but not necessary for payments in isolation: maintaining account balances is sufficient.    
\emph{Payments by themselves} are rarely interesting, however.  They are almost \emph{always} part of a \emph{compound} transaction: a spot exchange of a good or service for money; a life insurance or pension contract providing a complex interchange of moneys spread out over 50 years; or anything in between.

Complex transactions are particular \emph{sequences} of actions and events, \emph{some} of which (but not all) are payments.  
Enforcing correct execution of such sequences benefits from tamper-proof, nonrepudiable recording of sequences of events, including individual payments, not just updated account balances and guaranteeing that resources are not double-spent.  It is what blockchain and, more generally, distributed ledger systems, are about.  Blockchain as a trusted decentralized, persistent and tamper-proof resource storage and event recording layer that provides
the basis for contracts (protocols specifying obligations, permissions and prohibitions) amongst multiple parties and their execution strategies (``software-defined'' processes).

A fundamental challenge is linking physical assets to their digital representation, which requires some form of evidence that the digital representation of events recorded is \emph{correct}: Somebody could record in the blockchain that a physical asset, say a bicycle, has been delivered without it actually having been delivered in the real world.

Digital cash turns cash into a \emph{digital} asset: there is no distinction between the physical asset and its digital representation. Thus it eliminates the indirection between physical asset and its digital representation and the risk of inconsistency between the two. 

\section{Project topics}

The project is divided into 2 phases.  The deliverable of phase 1 is a preparatory analysis of the topics below, which is to serve as the basis for formulating an action plan for a focused project in phase 2.
 
\subsection{Dimensions of digital cash and its benefits and risks}

\begin{itemize}
\item  Traceable and non-traceable payments (KYC/AML, corruption resistance, desired levels of privacy);
\item  macro-economic and systemic consequences (e.g.~conversion of bank money to central bank money, evaluation of blockchain-hosted digital cash/stablecoins versus conventional banking payments, impact on tax collection and tax system);
\item physical and digital methods of controlling digital cash and their usability (physical tokens/cards including piggybacking on existing credit card, multi-factor authorization and authentication, prepaid cards, software-defined accounts) 
\item  potential of digital payments as part of contracts (complex transactions, cash for cash, cash for non-cash; tracking usage, automatic execution of compensating cash transactions upon contract failure), 
\item  recording and checking digital evidence for physical/non-cash transactions, 
\item  potential of noncollateralized (``promise'') contracts,
\item  potential of collateralized financial contracts (simple escrow, trade finance, etc.) and disintermediation by robotic execution of joint strategy (a.k.a.~smart contracts).
\end{itemize}

\subsection{Distributed systems architectures}

\begin{itemize}
\item taxonomy of centralized and distributed architectures, fundamental trade-offs 
\item identity management (cryptographic techniques for authentication and authorization) 
\item wallet/account management (multi-signature protocols, bank-managed digital wallets)
\item existing systems (Bitcoin, Ethereum, Corda, Hyperledger Fabric, ZCash etc.)
\item scalable distributed consensus (Sybil-attack resistant protocols for permissionless blockchain systems, crash- and Byzantine fault-tolerant protocols for permissioned blockchain systems)
\item P2P storage architectures (replicated state machines, structured P2P-systems, hierarchical distributed systems; conflict-free/eventually consistent replicated data types; probabilistic, partially consistency, amount- and geography-sensitive consistency) 
\item compositional blockchains (transferring assets/resources between different blockchains)
\end{itemize}

\subsection{Privacy and security}
\begin{itemize}
\item trusted computing platforms, taxonomy and consequences for privacy and security
\item cryptographic techniques
\begin{itemize}
\item PKI, symmetric keys, hashing, signatures, simple crypto protocols
\item zero knowledge proof
\item dynamic generation of keys and key-pairs for decorrelation
\end{itemize}
\item language-based techniques
\begin{itemize}
\item protocol (formal contract) design and analysis
\item mechanism design and analysis
\item static analysis (of self-executing contracts)
\item systems security techniques
\item formal proofs of correct implementations
\item attack vector analysis
\end{itemize}
\end{itemize}

\subsection{Economic and legal aspects}

\begin{itemize}
\item  Economic potential and consequences of blockchain-based solutions and digital cash for the Danish economy, specifically the financial sector;
\item  full reserve versus fractional reserve banking versus shadow money systems (non-fiat cryptocurrencies);
\item  potential of real-time systemic analysis of financial and other sectors;
\item  novel possibilities due to digitizing of cash (helicopter money, universal basic income, interest on digital cash, payment tracking, attacking contracts);
\item  potential of smart contract management derived from PSD2 on top of conventional payment infrastructure;
\item  compliance with GDPR and banking regulation.
\end{itemize}

\subsection{Financial use cases}

\begin{itemize}
\item  Packaging/tokenization of factoring (credit given until payment) for efficient refinancing for SMEs
\item  Financial contract management (from loans to structured products and OTC contracts), clearing and settlement
\item  Decentralized organizations and funding, ICOs
\item  VAT (guaranteeing that VAT paid is deducted at most once), dividend tax (guaranteeing that dividend tax refunds are claimed at most once)
\item  KYC/AML (payment tracking, checking what payments are for by investigating the contracts they are part of) 
\item  Simple escrow (money for delivery of goods)
\item  International money transfers
\item  Currency funds (smart tokens)
\item  Smart cash (software-defined digital cash accounts)
\end{itemize}

\end{document}